\documentclass[%
 reprint,
showkeys,
amsmath,amssymb,
aps,
]{revtex4-1}

\usepackage[utf8]{inputenc}
\usepackage[T1]{fontenc}

\usepackage[dvipsnames,table]{xcolor}
\colorlet{PerceptionColor}{RoyalPurple}
\colorlet{CognitionColor}{RedViolet}
\colorlet{LocomotionColor}{Gray}
\colorlet{AgentSelectedColor}{red}
\colorlet{SwapCandidateColor}{orange}
\definecolor{TargetOrientedColor}{RGB}{76,114,202}
\definecolor{CooperativeColor}{RGB}{0,153,0}

\usepackage{amsmath,amssymb}
\usepackage[english]{babel}
\usepackage{booktabs}
\usepackage[justification=centering]{caption}
\usepackage{float}
\usepackage{csquotes}
\usepackage{graphicx}
\graphicspath{{./Figures/}}
\usepackage{dcolumn}
\usepackage{bm}
\usepackage{hyperref}
\usepackage{listings}
\usepackage{multirow}
\usepackage{siunitx}
\usepackage{subcaption}
\usepackage{url}

\newcommand{\Ie}{That is}
\newcommand{\ie}{that is} 
\newcommand{\eg}{e.\,g.}
\newcommand{\code}[1]{\texttt{#1}}
\newcommand{\program}[1]{\texttt{#1}}

\newcommand{\sref}[1]{Sec.\,\ref{#1}}
\newcommand{\fref}[1]{Fig.\,\ref{#1}}

\newcommand{\tref}[1]{Tab.\,\ref{#1}}

\newcommand{\lref}[1]{List.\,\ref{#1}}

\newcommand{\ii}[1]{\textbf{(#1)}}
\newcommand{\iih}[2]{\textbf{(#1) #2:}}

\newcommand{\mean}[1]{\bar{#1}}

\newcommand{\norm}[2][2]{||#2||_{#1}}
\newcommand{\realnumbers}{\mathbb{R}}

\providecommand{\gertaaccepted}[1]{#1}

\providecommand{\johnaccepted}[1]{#1}

\begin{document}



\title{Agent-Based Simulation of Collective Cooperation: From Experiment to Model}
\thanks{A footnote to the article title}%

\author{Benedikt Kleinmeier}
\email{benedikt.kleinmeier@hm.edu}
\affiliation{Munich University of Applied Sciences, Department of Computer Science and Mathematics, 80335 Munich, Germany}
\affiliation{Technical University of Munich, Department of Informatics, 85748 Garching, Germany}

\author{Gerta Köster}%
\email{gerta.koester@hm.edu}
\affiliation{%
 Munich University of Applied Sciences, Department of Computer Science and Mathematics, 80335 Munich, Germany
}

\author{John Drury}%
\email{j.drury@sussex.ac.uk}
\affiliation{%
University of Sussex, School of Psychology, BN1 9RH, Brighton, United Kingdom
}


\date{\today}

\begin{abstract}
\gertaaccepted{Simulation models of pedestrian dynamics have become an invaluable tool for evacuation planning. Typically crowds are assumed to stream unidirectionally towards a safe area. Simulated agents avoid collisions through mechanisms that belong to each individual, such as  
being repelled from each other by imaginary forces. But classic locomotion models fail when collective cooperation is called for, notably when an agent, say a first-aid attendant, needs to forge a path through a densely packed group.
We present a controlled experiment to observe what happens when humans pass through a dense static crowd. We formulate and test hypothesis on salient phenomena. 
We discuss our observations in a psychological framework. We derive a model that incorporates: agents' perception and cognitive processing  of a situation that needs cooperation; selection from a portfolio of  behaviours, such as being cooperative; and a suitable action, such as swapping places. 
Agents' ability to successfully get through a dense crowd emerges as an effect of the psychological model. 
}
\end{abstract}

\pacs{Valid PACS appear here}
\maketitle



\section*{Introduction}
\label{sec:Introduction}

Simulation models of pedestrian dynamics are widely used today especially for evacuation planning \cite{helbing-2001,schadschneider-2009,moussaid-2011,viswanathan-2014}. Such models usually consist of unidirectional flows of agents (simulated pedestrians) and are used to estimate the evacuation time in emergency situations or to test safety concepts \cite{fruin-1993}. Simulations of such models are a useful tool in the planning phase to detect critical high densities for example to avoid casualties like reported at the Hajj several times \cite[p. 164]{challenger-2009} or at the Love Parade music festival 2010 in Germany \cite{helbing-2012}.

\begin{figure}[!h]
    \begin{subfigure}{0.30\linewidth}
        \includegraphics[width=\linewidth]{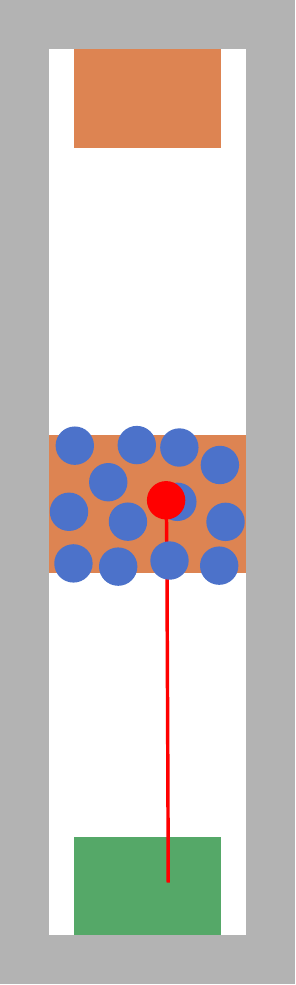}
        \caption{Social Force Model}
    \end{subfigure}
    \begin{subfigure}{0.30\linewidth}
        \includegraphics[width=\linewidth]{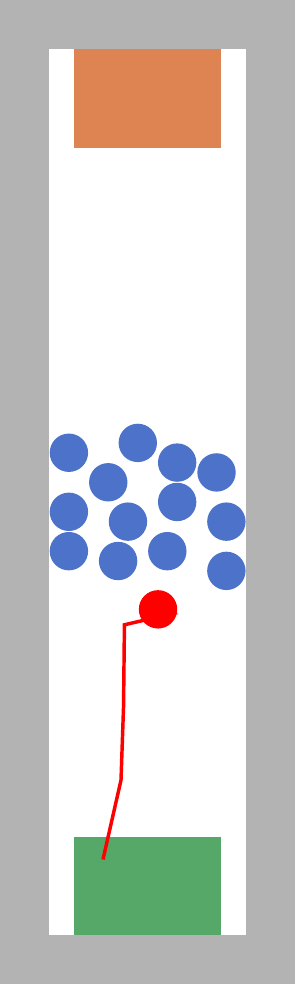}
        \caption{Optimal Steps Model}
    \end{subfigure}
    \caption{A walking agent (red) starts walking at the bottom area and tries to reach the rectangular target area on top while confronted with a dense, stationary crowd. In simulations, we cannot identify real interaction between agents when using different physically-inspired locomotion models. Either the walking agent \enquote{ignores} the dense, stationary crowd and walks on other agents which could not happen in real life (a). Or the crowd blocks the walking agent completely because of the high density (b). The open source simulator \program{Vadere} was used for the simulations.}
    \label{fig:SimulationsAndBlockedAgent}
\end{figure}

Locomotion models \cite{gipps-1985,helbing-1995,seitz-2012} work well for unidirectional flows because they are mostly validated against empirical data \cite{haghani-2018}. They can provide helpful insights and make crowd gatherings safer. \gertaaccepted{But often locomotion models fail for setups that seem only slightly different.} For instance, when a first aid-attendant needs to forge a path through a dense crowd to reach an injured person. When reenacting such a real-world situation in current simulation tools, agents often get stuck and end up in a deadlock situation because there is no real interaction between agents, compare \fref{fig:SimulationsAndBlockedAgent}.

\subsection*{Related Work}
\label{sec:RelatedWork}

Several authors extended existing locomotion models to manoeuvre agents through virtual environments and to mitigate shortcomings of these models. For instance, \cite{moussaid-2009,smith-2009} let agents evade tangentially or sideways. Using such collision avoidance strategies on a microscopic level often leads to lane formation on a macroscopic level. But, pure physically inspired locomotion strategies like collision avoidance do not work for very dense crowds as seen by the simulations in \fref{fig:SimulationsAndBlockedAgent}. In the real world, humans adapt their behaviour \cite[p. 11--12]{gerrig-2013}. For instance, humans just ask to be let through. Humans use perception, cognition and a repertoire of different behaviours. \cite{pelechano-2007}, made first steps to integrate psychological findings into pedestrian dynamic simulations to control the Social Force Model \cite{helbing-1995}. \cite{pelechano-2007} integrated different agent states like queuing or pushing behaviour. \cite{yanagisawa-2016,xue-2020} extended cellular automata to overcome deadlock situations in bidirectional pedestrian flows based on a more cooperative behaviour of agents. \cite{yanagisawa-2016} integrated ideas from game theory and swerving preferences which are based on previous successful swerving behavior. \cite{xue-2020} conducted an own experiment \cite{xue-2017} and integrated a sort of \enquote{give way to counterflowing agents} into a cellular automaton. However, both approaches were limited to cellular automata models, \cite{yanagisawa-2016} even to one-dimensional scenarios. Also other simulator developers, both researchers and commercial ones \cite{jupedsimcontributors-2015,accurate-2020}, extended existing locomotion models to better cover also waiting behaviour and other real-world situations. But all these extensions were integrated without providing empirical data or evidence. In contrast, \cite{wijermans-2013} provided empirical evidence and developed a model to simulate crowd behaviour with social-cognitive agents with a focus on music festivals. They captured the motivation of individuals by including various physiological parameters like memory, bladder, stomach and arousal and goal-oriented agents. But adding a plethora of parameters on the individual level makes the model difficult to understand. \cite{feliciani-2016} extended a cellular automaton to allow greater densities and enabled swapping strategies for agents to maintain flow in counterflow scenarios. Nevertheless, to our knowledge there is no systematic operationalisation of psychological processes to let agents pass through a stationary crowd. We argue that the classic locomotion models don't capture collective cooperation of real humans.

For us, the classic modelling process consists of making real-world observations and then finding mathematical and algorithmic formulations to describe the observed phenomena well. After implementing this as computer programs, we are able to carry out simulations to get further insights. In fact, another reason why models for high-density situations are still missing is the lack of empirical data. Numerous authors conducted experiments with a strong focus on unidirectional flow of pedestrians with moderate density \cite{helbing-2005,zhang-2011,zhang-2014,templeton-2018} and counterflow scenarios \cite{zhang-2012b,templeton-2019} or bottlenecks \cite{daamen-2003,seyfried-2009}. Other authors focused more on collective phenomena in crowds. For instance, \cite{sieben-2017} investigated the influence of barriers on the behaviour of participants. They included the social psychology perspective by using questionnaires to get insights into participants' perception. And other authors focused more on egress and queuing behaviour like \cite{wagoum-2017}. To our knowledge, \cite{nicolas-2019} are the first authors who conducted an experiment with a stationary crowd and who tested the effects on walking participants. Even the exhaustive two-volume literature review \cite{haghani-2020a,haghani-2020b} for empirical methods and experiments in pedestrian dynamics did not explicitly mention stationary crowds and their effects. So far, often an experiment was conducted but no model derived, or the model stopped at a verbal description while a mapping to a clean and reusable software architecture is missing. \gertaaccepted{In fact, until recently, simulation frameworks for pedestrian dynamics completely lacked evidence-based models of cooperative actions, such as group actions \cite{templeton-2015}. Since then, first proofs of concept of specific situations have emerged, where empirical findings from social psychology, not analogy from physics, inspire the model. See  \cite{sivers-2014,sivers-2016d}. Yet, to our knowledge, nobody has operationalised psychological findings into computer models of crowds where an agents' ability to pass through a dense crowd emerges as an effect.} We would like to close this gap.

\gertaaccepted{But what should be the corner stone of such a psychological model?} \johnaccepted{Prima facie, it seems that individuals manage to flow through dense crowds, and that this is achieved via cooperation from the crowd, who adjust themselves and move to give the individual a little space, rather than via force (since the latter would breach social norms around peaceful behaviour and politeness). For example, \cite{templeton-2018} show that when individuals approach a crowd in counterflow they do not simply walk into it nor do they simply stop but rather there is some negotiation of space among individuals to allow one to flow through the other.} 
\gertaaccepted{Thus, we argue that we need a model of crowd cooperation.}

\subsection*{Goals of our Work and Article Structure}
\label{sec:GoalsOfOurWorkAndArticleStructure}

\gertaaccepted{In this contribution, we aim to model collective behaviour in a crowd so that the ability of agents to pass through dense static crowds emerges. Our goal is to directly base the model on empirical evidence and also to firmly put it into the frame of current social psychology. Finally, we strive for a reusable software structure and free and open-source implementation of the model that can be generalized to a large number of instances of cooperative collective behaviour.}

The paper is structured as follows:
In \sref{sec:Experiment}, we present a controlled experiment which we conducted in 2018 with students. We list our observations and formulate hypothesis which we test statistically. The most important, albeit almost trivial observation is, that the participants were indeed able to get through the crowd. 
In \sref{sec:Model}, we then describe the psychological processes in such a situation and operationalise it into a parsimonious model. That is, we restrict the model to elements that we deem absolutely necessary: the perception and subsequent cognition of a situation that calls for behavioural changes, the selection of a behaviour from a portfolio, notably being cooperative, and the selection of a suitable action, such as swapping positions. This operationalisation represents a simple, generic and reusable model allowing more interactions between agents which can be easily implemented in different pedestrian simulation programs.
We implement our model in a parsimonious computer program for which we run computer experiments in which the desired phenomenon emerges: Agents are able to pass through a crowd. Finally, in  \sref{sec:Conclusion}, we evaluate our results. We present ideas how to add detail for a better quantitative match of second order effects and we discuss how to generalize the model to encompass other collective phenomena.

\section{Experiment}
\label{sec:Experiment}

\subsection{Experiment Set-Up}
\label{sec:PresentationSetUp}

In order to study the effects of high densities on a walking person, we performed a controlled experiment in the foyer of the Munich University of Applied Sciences on Oct 12, 2018 (11:45 -- 13:00).

In the experiment, we observed how a participant walks through a dense, waiting crowd. To this end, we kept 58 participants in a separate waiting room. The participants were entertained by experiment assistants with quizzes and discussions to keep the atmosphere as normal as possible. To avoid any priming, the participants received minimal information. The participants signed an informed consent form with the title \enquote{Study on movements of pedestrians}. The form stated that no physical risks were involved and that the experiment was recorded on camera. We chose first-year students in their second week as participants to ensure that they did not know anything about the experiment's intentions.

\begin{figure}[!h]
    \centering
    \begin{subfigure}{0.45\linewidth}
        \centering
        \includegraphics[height=\linewidth]{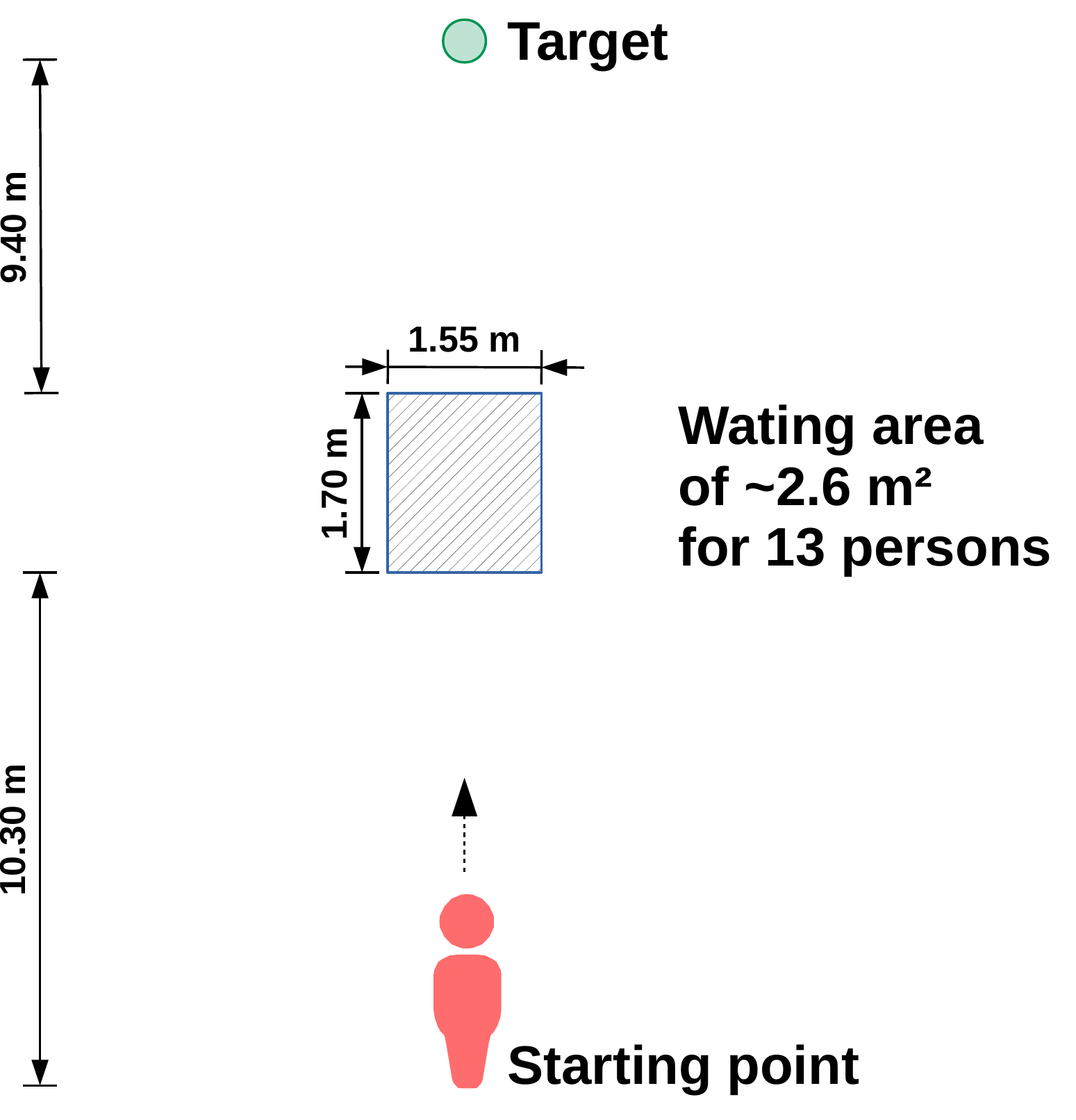}
        \caption{Schematic setup}
        \label{fig:ExperimentDenseCrowdSchematicSetup}
    \end{subfigure}
    \begin{subfigure}{0.45\linewidth}
        \centering
        \includegraphics[height=\linewidth]{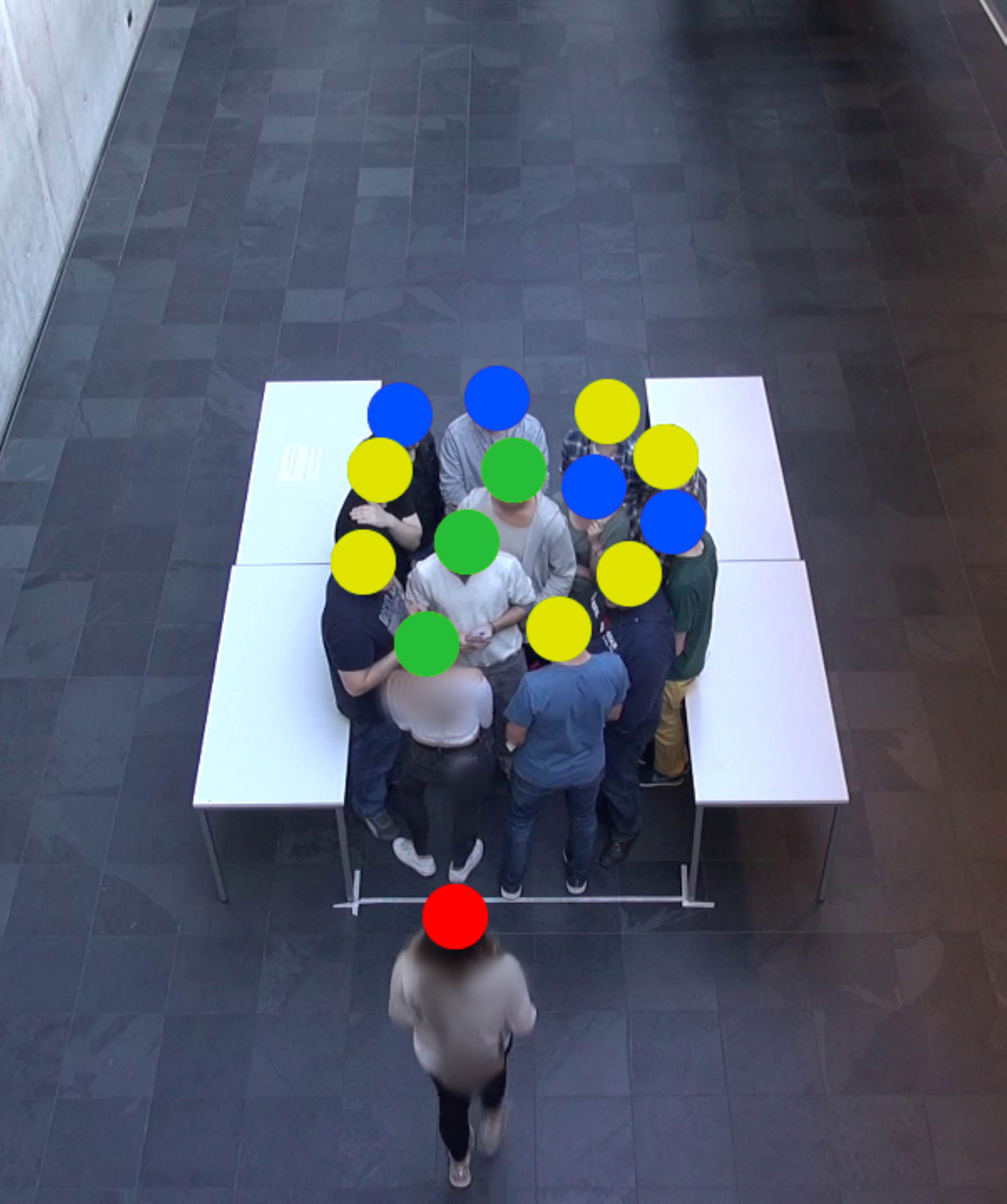}
        \caption{Real setup}
        \label{fig:ExperimentDenseCrowdRealSetup}
    \end{subfigure}
    \caption{The experiment setup: A waiting crowd of 13 participants in a delimited area of \SI{2.64}{\meter\squared} is successively crossed by a participant (figure from \cite{kleinmeier-2019b}).}
    \label{fig:ExperimentDenseCrowd}
\end{figure}

During the experiment, 13 participants stood in a delimited area of \SI{2.64}{\meter\squared} (\SI{1.55}{\meter} \(\times\) \SI{1.70}{\meter}) as a waiting crowd. In each experiment run, the walking participant successfully crossed the crowd along what would be the y-axis in \fref{fig:ExperimentDenseCrowdSchematicSetup}. The density while crossing was \(\rho = \SI{5.30}{ped\per\meter\squared}\). For each experiment run, we randomly chose one person from the waiting room and assigned this person as walking participant. For the very first run, we also chose 13 persons from the waiting room and assigned them as waiting crowd.

We took two measures to avoid training effects for the waiting crowd: \ii{1} After each run, a staff member shuffled the waiting crowd. To this end, the waiting crowd were asked to leave and re-enter the waiting area, so that the positions of the participants were shuffled. \ii{2} After five runs, seven random participants of the waiting crowd were replaced by seven participants from the waiting room, who were also chosen randomly. We also took several measures to avoid observer biases like using a standardized experiment procedure with consistent instructions for all participants. The walking participants were instructed with the sentence \enquote{Go to the tree by crossing the crowd}. The waiting crowd was instructed with \enquote{Wait in the delimited area}. See \cite{kleinmeier-2019b} for a description of all measures. Tables on the left- and right-hand side of the waiting area prevented the participants from leaving the waiting area accidentally.


The experiment set-up is depicted in \fref{fig:ExperimentDenseCrowd} and described in more detail in \cite{kleinmeier-2019b}.

\begin{figure}[!h]
    \begin{subfigure}{0.45\linewidth}
        \includegraphics[width=\linewidth]{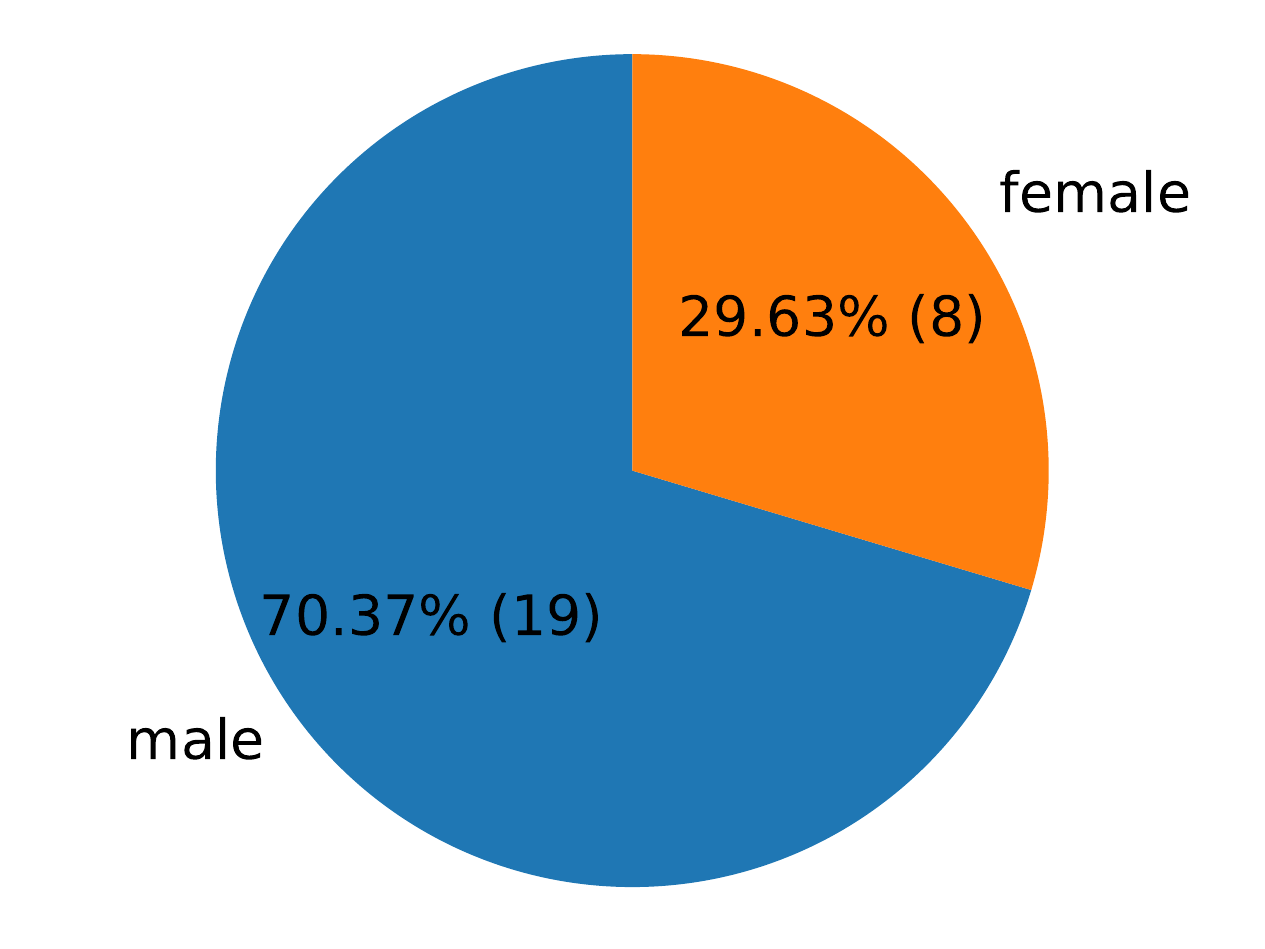}
        \caption{Gender}
    \end{subfigure}
    \begin{subfigure}{0.45\linewidth}
        \includegraphics[width=\linewidth]{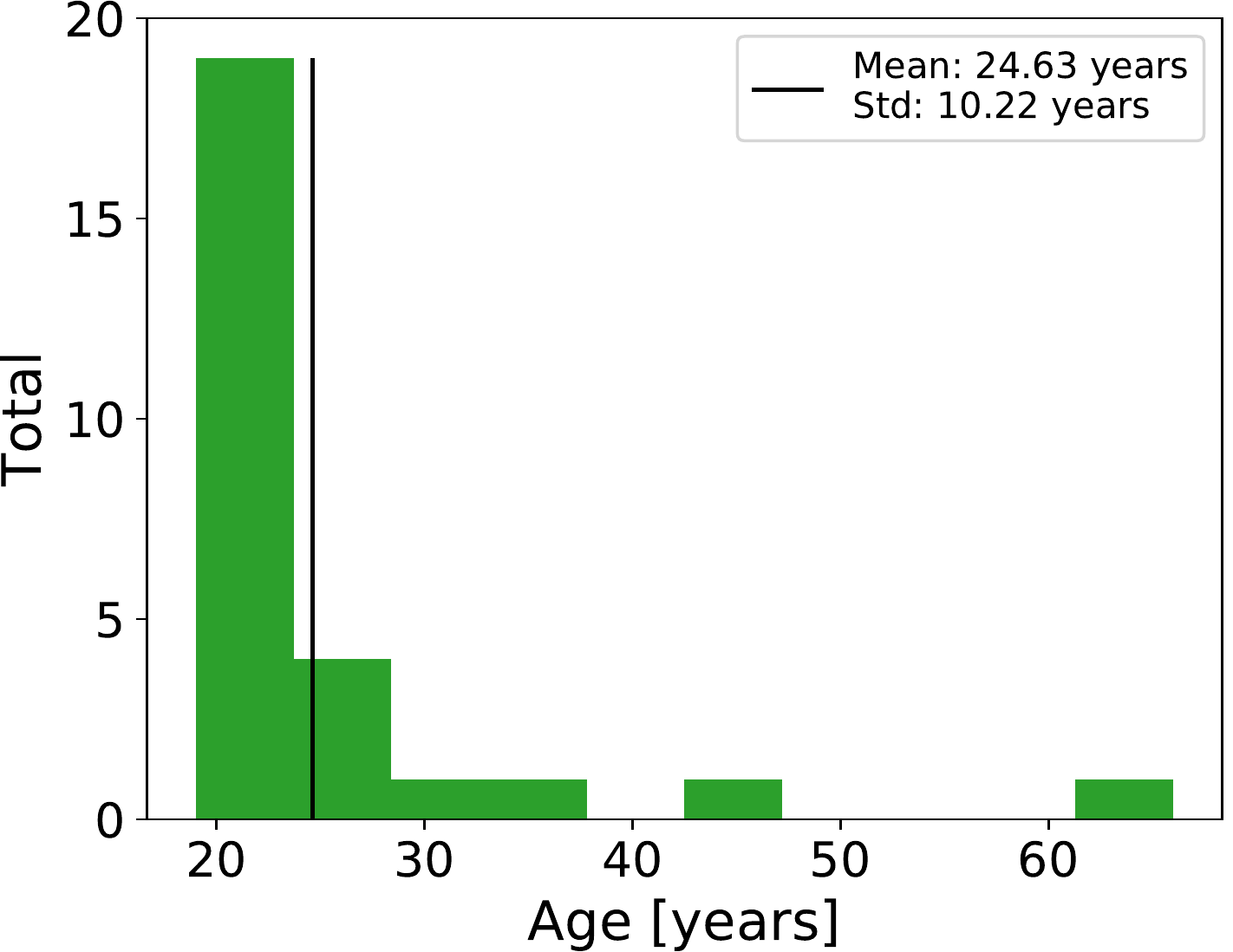}
        \caption{Age}
    \end{subfigure}
    \caption{Gender and age distribution of the 27 walking experiment participants.}
    \label{fig:ExperimentDenseCrowdTestPersons}
\end{figure}

In total 58 students participated in the experiment. 27 of them, (men and women), aged 19--66, were assigned as walking participants and performed 30 runs (compare \fref{fig:ExperimentDenseCrowdTestPersons}). We collected gender, age, height and shoulder width for each walking participant.

The experiment was filmed from above at an angle of around \ang{60} (compare \fref{fig:ExperimentDenseCrowdRealSetup}). We recorded the experiment with a camcorder Sony Handycam HDR-PJ780VE using a resolution of \SI{1280}{pixel}\(\times\)\SI{720}{pixel} and \SI{25}{frames\ per\ second}. The raw video material had a length of \SI{73}{minutes}. We used the free video analysis and modelling tool \program{Tracker} \cite{trackercontributors-2019} to correct the optical distortion and to track the trajectories of the walking participant and the waiting crowd. For this purpose, we applied \program{Tracker's} \enquote{Auto-Tracker} feature. See section \ref{sec:TrajectoryExtraction} for more information about trajectory extraction. After trajectory extraction, we used self-written Python scripts, more precisely Jupyter notebooks, to analyse the data.

\subsection{Experiment Results}
\label{sec:ExperimentResults}

We began by watching the experiment's video footage. This step helped us to verbalize the human behaviour we observed and to formulate the following hypotheses:

\begin{itemize}
    \item Pedestrians walking through a crowd are slowed down.  
    \item The pedestrians in a waiting crowd return to their initial positions after giving way to the \enquote{intruder}.
    \item Real humans can pass a crowd at high densities.  
\end{itemize}

The last hypothesis, while seemingly trivial, is the most important one, because this is where simulated agents have failed so far. In a second step we will test these hypotheses and quantify effects.  

\subsubsection{Experiment Result: Speed Distributions}
\label{sec:ExperimentResultSpeedDistributions}

Firstly, we measured the instantaneous speed of the walking participant inside and outside the waiting crowd. Outside the crowd this gives an estimate of the \enquote{free-flow} speed, which is the walking speed of a pedestrian if no external effects force the pedestrian to slow down or to speed up. We measured the \enquote{free-flow} speed in front of the waiting crowd instead of behind because the area in front is closer to the camera and we expect a lower measurement error from optical distortion. The instantaneous speed \(v_{i}(t)\) for walking participant \(i\) at time step \(t\) is defined as:

\begin{equation}
v_{i}(t) = \frac{ \sqrt{{\Delta x_{i}(t)}^2 + {\Delta y_{i}(t)}^2} }{ \Delta T }
\end{equation}

where \(\Delta T = \SI{1/25}{\second} = \SI{0.04}{\second}\) (\ie, in the \program{Tracker} software we evaluated 25 camera frames per second) and \({\Delta x_{i}(t)} = x_{i}(t) - x_{i}(t - 1)\).

Then we averaged the instantaneous speed values \(v_{i}(t)\) over all time steps \(N\) when the participant was inside the measurement area:

\begin{equation}
\mean{v}_{i} = \frac{ 1 }{ N } \sum_{t=1}^{N} v_{i(t)}
\end{equation}

\begin{figure}[!h]
    \centering
    \includegraphics[width=\linewidth]{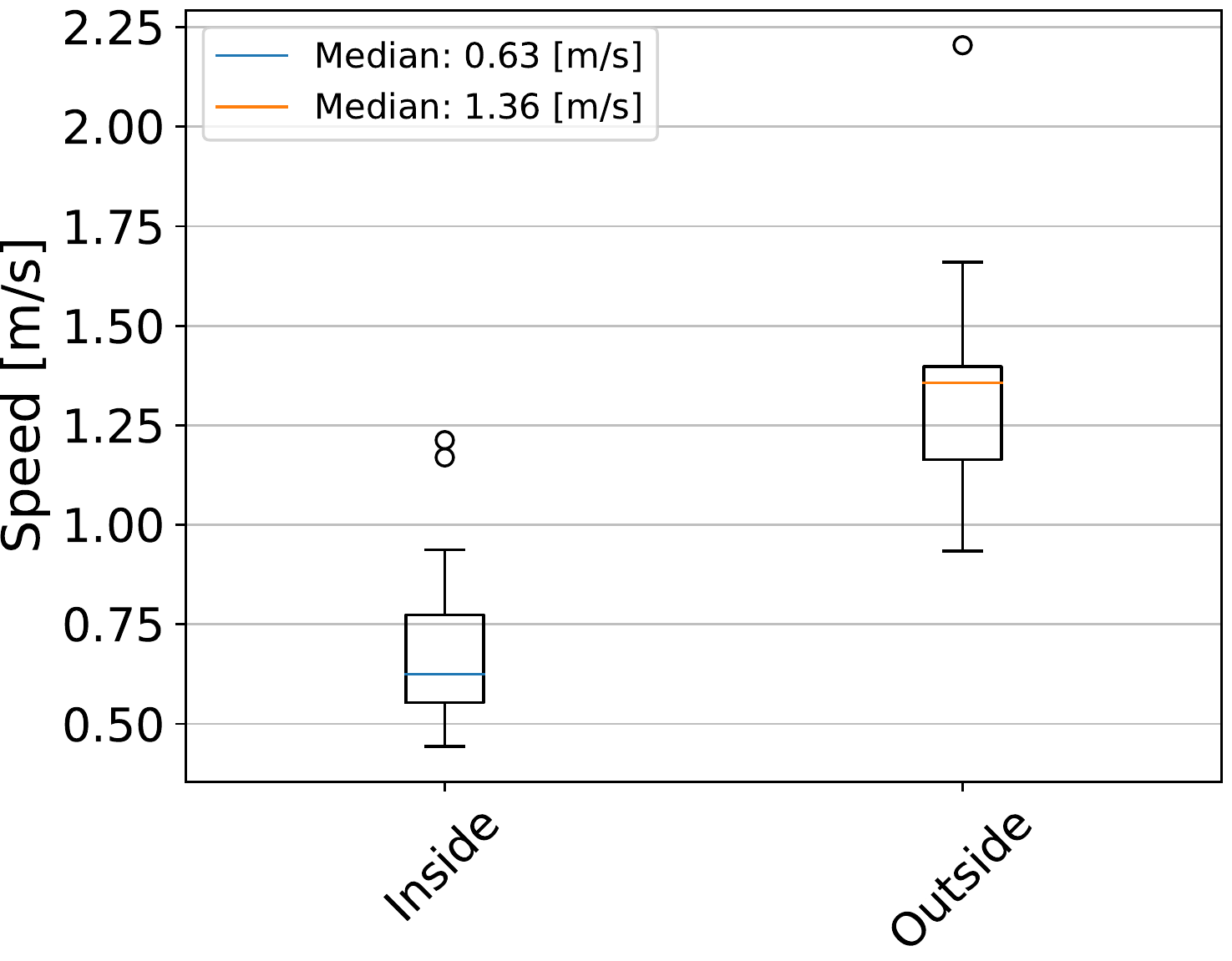}
    \caption{Box plot for speed distribution (averaged instantaneous speeds) of the walking participants inside and outside the waiting crowd}
    \label{fig:ExperimentDenseCrowdSpeedBoxplot}
\end{figure}

\fref{fig:ExperimentDenseCrowdSpeedBoxplot} and \tref{tbl:ExperimentDenseCrowdSpeedStatistic} provide an overview of the averaged instantaneous speeds of all walking participants.

\begin{table}[!h]
    \begin{tabular}{lrr}
        \toprule
        {} &         \multicolumn{2}{c}{\textbf{Speed [\si{\meter\per\second}]}} \\
        {} &         \textbf{Inside} &  \textbf{Outside} \\
        \midrule
        sample size & 30.00 &      30.00 \\
        mean  &  0.70 &       1.33 \\
        std   &  0.19 &       0.25 \\
        min   &  0.44 &       0.93 \\
        25\%   &  0.55 &       1.16 \\
        50\%   &  0.63 &       1.36 \\
        75\%   &  0.77 &       1.40 \\
        max   &  1.21 &       2.20 \\
        \bottomrule
    \end{tabular}
    \caption{Detailed statistics for the measured speed distributions of the walking participants inside and outside the waiting crowd.}
    \label{tbl:ExperimentDenseCrowdSpeedStatistic}
\end{table}

Comparing the mean instantaneous speed of \SI{0.70}{\meter\per\second} (inside) and \SI{1.33}{\meter\per\second} supports our hypothesis that the walking participants are slowed down by the waiting crowd. At a density \(\rho\) of 14 persons per \SI{2.64}{\square\meter}, that is, \(\rho = \SI{5.30}{persons\per\square\meter}\), we measure a \enquote{slow-down} factor of \(\frac{ \SI{1.33}{\meter\per\second} }{ \SI{0.70}{\meter\per\second} }
= 1.9 \approx 2 \).

We also performed  Student's t test to check if the waiting crowd has an effect on a walking participant's speed. For this, we calculated for each walking participant \(i\): \(\Delta v_{i} = v_{i,in} - v_{i,out}\). Then, we applied a one-sided t-test with following mathematical hypotheses:

\begin{itemize}
    \item \(H_{0}: mean(\Delta v_{i}) \geq 0\)
    \item \(H_{1}: mean(\Delta v_{i})  < 0\)
    \item Significance level: 0.05
\end{itemize}

The test statistic \( T = \sqrt{N} \times \frac{mean(\Delta v_{i}) - 0}{std(\Delta v_{i})} \) revealed a value of \(T = -10.75\) for all \(N=30\) participants. We drop the \(H_{0}\) hypothesis of no influence since our tests statistic \(T\) is far below the significance limit of 0.05 of the corresponding t distribution, which is \(-1.70\) at a p-value~\(\ll 0.01\). 

When watching the video footage, we identified some potential outliers in the data. For instance, we observed a particular fast participant outside and inside the waiting crowd. The participant stretched out the hands like a swimmer to \enquote{dive} through the crowd. We also observed a very slow participant inside the waiting crowd whom some members of the waiting crowd  blocked intentionally. We decided to keep these outliers for our statistical analysis  to stay close to the real world where one can also observe different techniques to cross a dense crowd. Some of these techniques are faster or slower than others.

The measured mean free-flow velocity of \SI{1.33}{\meter\per\second} outside the waiting crowd is very close to previous empirical measurements like \cite{weidmann-1992} with \SI{1.34}{\meter\per\second}. This strengthens our belief that we gathered realistic data.

\subsubsection{Experiment Result: Distribution of Waiting Crowd}
\label{sec:ExperimentResultDistribution of Waiting Crowd}

We want to shed light on the question if participants of the waiting crowd return to their initial positions after giving way to an intruder. To analyse the movement of each participant of the waiting crowd, we looked at two metrics: First, we measured the Euclidean distance between the initial and the end position of each participant. Second, we looked at the maximum Euclidean distance a waiting participant walked. For this, we compared each position of a participant's trajectory\footnote{A trajectory encapsulates the positions of a participant over time.} with the trajectory's initial position.

\begin{figure}[!h]
    \centering
    \includegraphics[width=\linewidth]{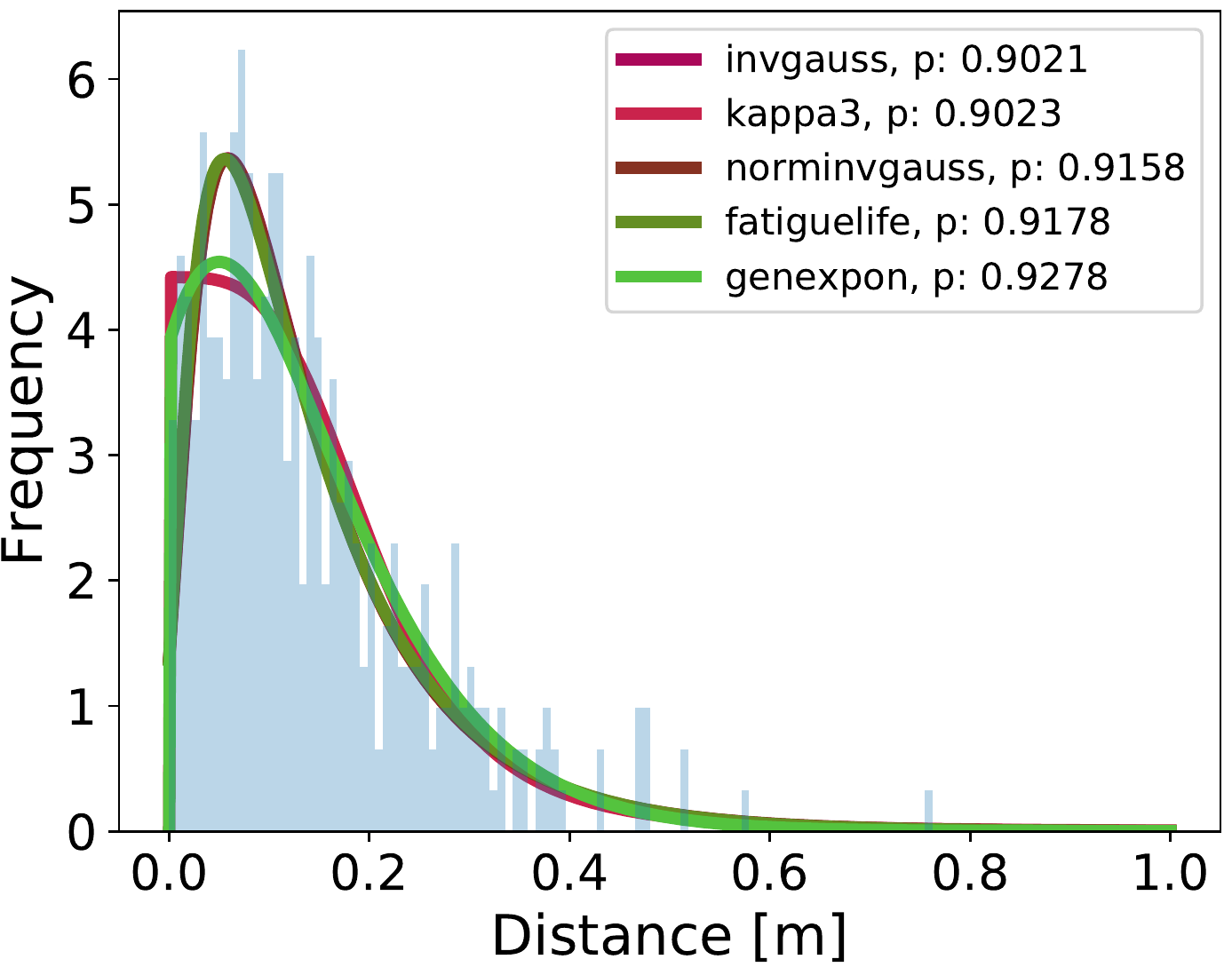}
    \caption{The data in blue visualizes the Euclidean distances between a participant's initial position --- before the walking participant entered the waiting crowd --- and the end position. The Euclidean distance is defined as \(\norm[2]{p_{initial} - p_{end}}\) with \(p \in \realnumbers^{2}\). The plot includes the best-fitting continuous distributions with a p-value \(\ge 0.90\).}
    \label{fig:ExperimentDenseCrowdWaitingCrowdStartEndMetricBestDistributions}
\end{figure}

\begin{figure}[!h]
    \centering
    \includegraphics[width=\linewidth]{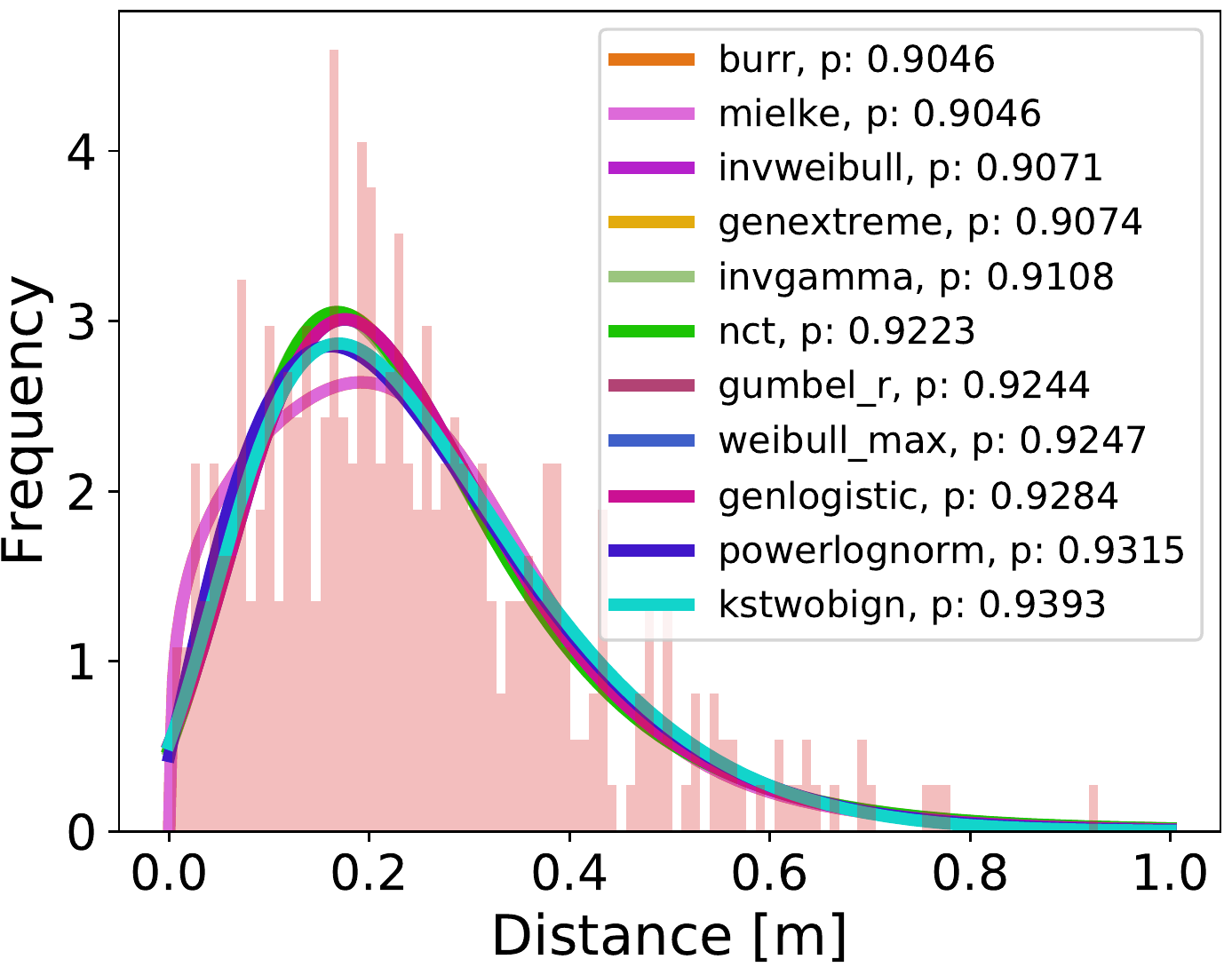}
    \caption{The data in red visualizes the maximum Euclidean distance a participant of the waiting crowd moved while the crowd was crossed by the walking participant. The plot includes the best-fitting continuous distributions with a p-value \(\ge 0.90\).}
    \label{fig:ExperimentDenseCrowdWaitingCrowdMaxDistanceMetricBestDistributions}
\end{figure}

Then, we investigated if the extracted distances follow a continuous probability distribution. We tested the data against 94 distributions\footnote{\url{https://docs.scipy.org/doc/scipy/reference/stats.html\#continuous-distributions}} and used the Kolmogorov-Smirnov test to verify the goodness of fit. The Kolmogorov-Smirnov test assumes as null hypothesis \(H_{0}\) that the sampled data and the tested distribution follow the same probability distribution. In our survey, we keep only distributions with a p-value greater 0.90.
\fref{fig:ExperimentDenseCrowdWaitingCrowdStartEndMetricBestDistributions} and \tref{tbl:ExperimentDenseCrowdWaitingCrowdStartEndMetricStatistic} summarize the data for the first metric (the Euclidean distance between initial and end position). \fref{fig:ExperimentDenseCrowdWaitingCrowdMaxDistanceMetricBestDistributions} and \tref{tbl:ExperimentDenseCrowdWaitingCrowdMaxDistanceMetricStatistic} summarize the data for the second metric (the maximum Euclidean distance).

\begin{table}[!h]
    \begin{tabular}{lr}
        \toprule
        {} &   \textbf{Distances [\si{\meter}]} \\
        {} &   \textbf{(Metric 1)}\\
        \midrule
        sample size &     400.00 \\
        mean  &       0.14 \\
        std   &       0.11 \\
        min   &       0.00 \\
        25\%   &       0.06 \\
        50\%   &       0.11 \\
        75\%   &       0.19 \\
        max   &       0.76 \\
        \bottomrule
    \end{tabular}
    \caption{Detailed statistics for the participants of the waiting crowd and the Euclidean distance between participant's initial and end position.}
    \label{tbl:ExperimentDenseCrowdWaitingCrowdStartEndMetricStatistic}
\end{table}

\begin{table}[!h]
    \begin{tabular}{lr}
        \toprule
        {} &   \textbf{Distances [\si{\meter}]} \\
        {} &   \textbf{(Metric 2)} \\
        \midrule
        sample size &     400.00 \\
        mean  &       0.25 \\
        std   &       0.16 \\
        min   &       0.00 \\
        25\%   &       0.13 \\
        50\%   &       0.22 \\
        75\%   &       0.33 \\
        max   &       0.93 \\
        \bottomrule
    \end{tabular}
    \caption{Detailed statistics for the participants of the waiting crowd and the maximum Euclidean distance}
    \label{tbl:ExperimentDenseCrowdWaitingCrowdMaxDistanceMetricStatistic}
\end{table}

\gertaaccepted{We cannot identify one single and best-fitting distribution for each of the two metrics. However, we observe that the best fitting distributions are not of the same type for the two metrics. We also observe that the distribution of the maximum distance is broader, with a heavy tail towards a larger value.

We hypothesized that participants in the waiting crowd return to their initial positions. But, it would be unrealistic to expect them to hit the exact same spot. Also, people shift from one foot to the other which causes the head to sway for at least several centimetres which is also reported by \cite[Fig.\,3, p. 4]{boltes-2016}. Thus, within an error margin, we would expect a distribution for the first metric, which is centreed around a value, a little off from zero, by which waiting individuals, on average, miss their original position.  This, in principle, is what we see. In any case, the participants do not stay at the position of maximum difference.}

We argue that the data supports a tendency to return, where the mean distance from the initial position is only \SI{0.14}{\meter} with a standard deviation of \SI{0.1}{\meter}.


\subsubsection{Experiment Result: Trajectories and Walking Participant and Duration in Waiting Area}
\label{sec:ExperimentResultTrajectoriesAndWalkingParticipantAndDurationInWaitingArea}

\begin{figure}[!h]
    \centering
    \includegraphics[width=\linewidth]{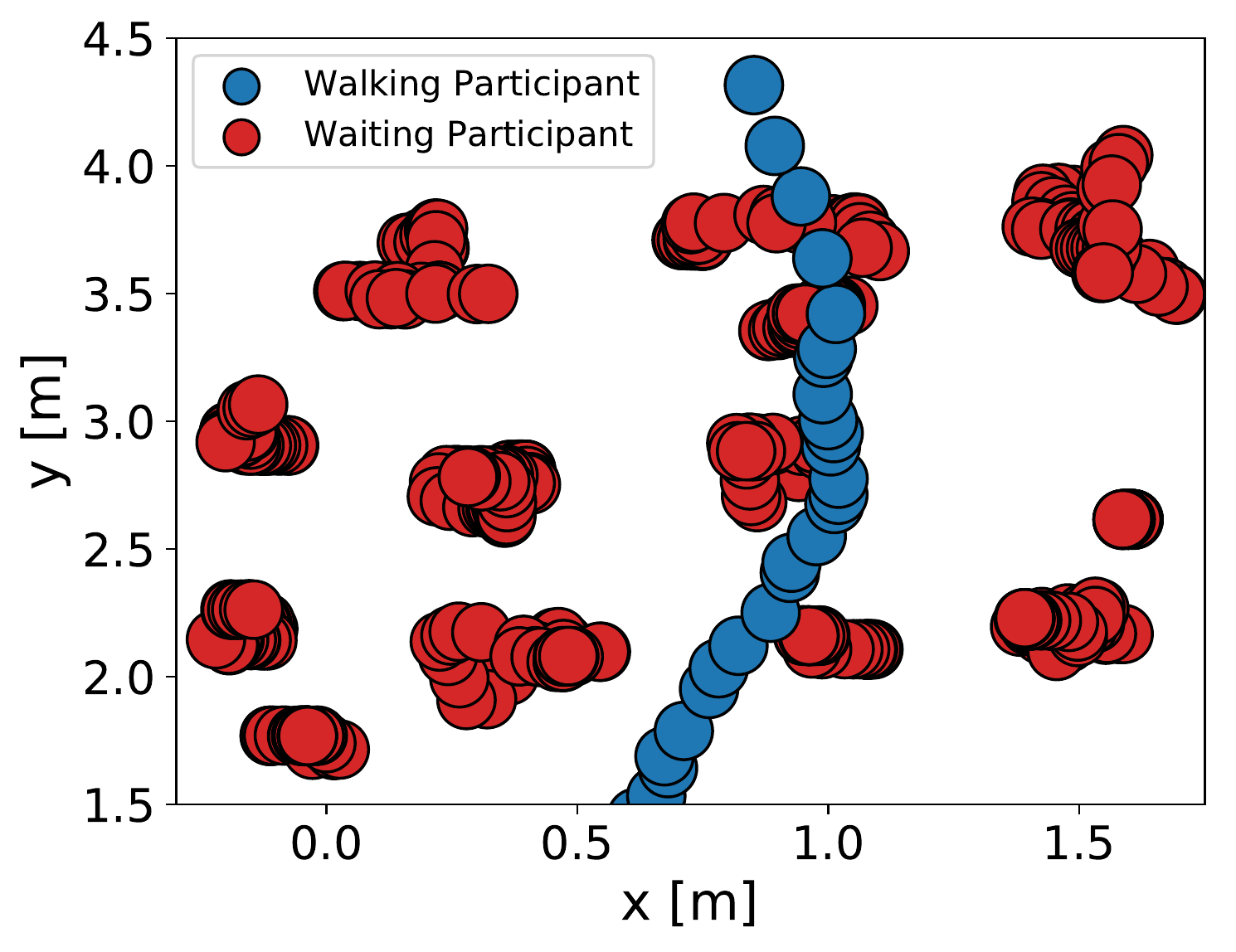}
    \caption{The trajectory of a single walking participant inside the waiting area at a time resolution of \SI{1/25}{\second}.}
    \label{fig:SingleTrajectoryTimeStep0_8}
\end{figure}

\begin{figure}[!h]
    \centering
    \includegraphics[width=\linewidth]{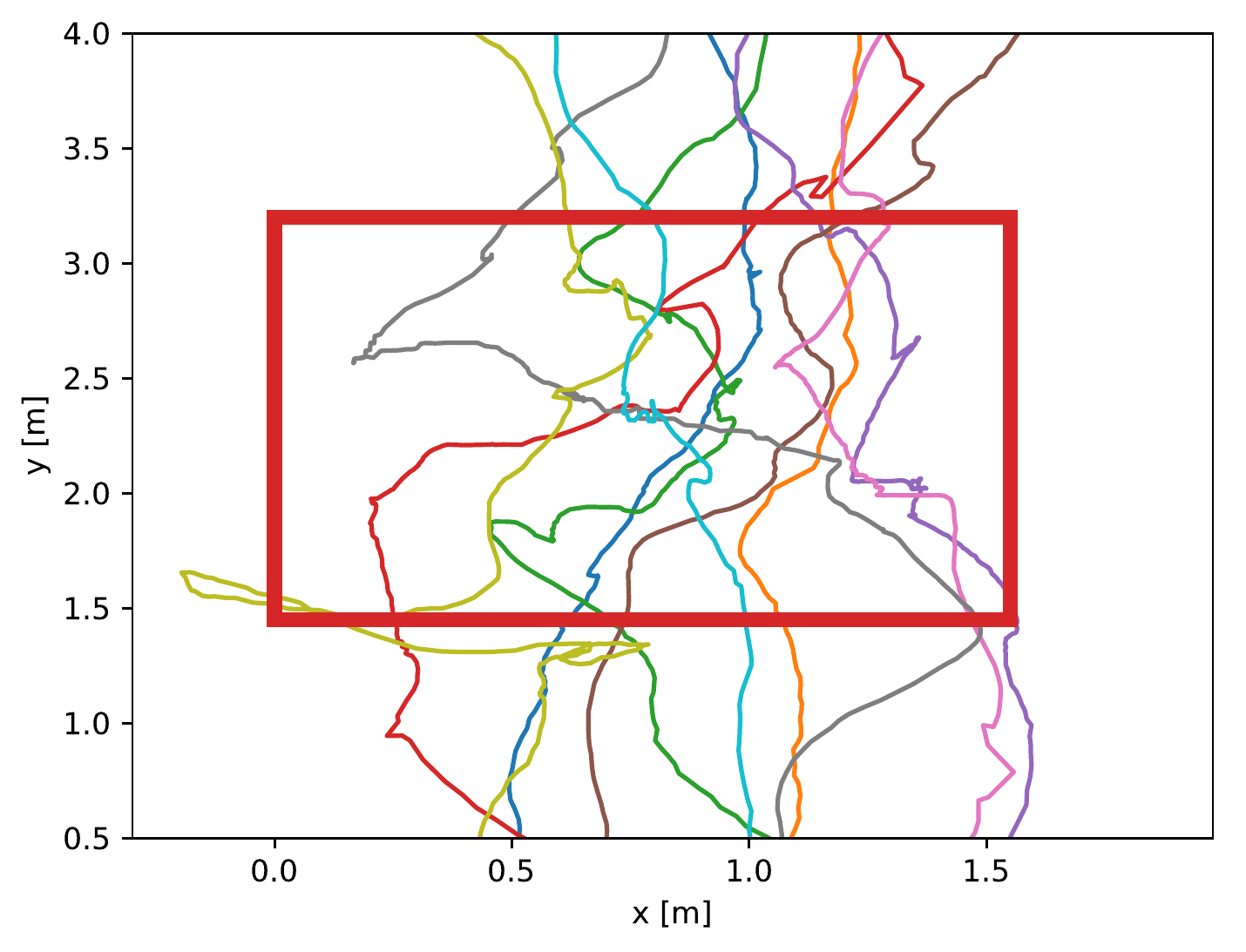}
    \caption{The trajectories of ten walking participants inside the waiting area (red rectangle) at a time resolution of \SI{1/25}{\second}.}
    \label{fig:MultipleTrajectoriesStepSize3}
\end{figure}

With a third set of measurements, we took a  closer look on how  walking participants manoeuvre through the waiting crowd. For this, we first plotted the trajectories of the walking participants, see \fref{fig:SingleTrajectoryTimeStep0_8} and \fref{fig:MultipleTrajectoriesStepSize3}.

\begin{figure}[!h]
    \centering
    \includegraphics[width=\linewidth]{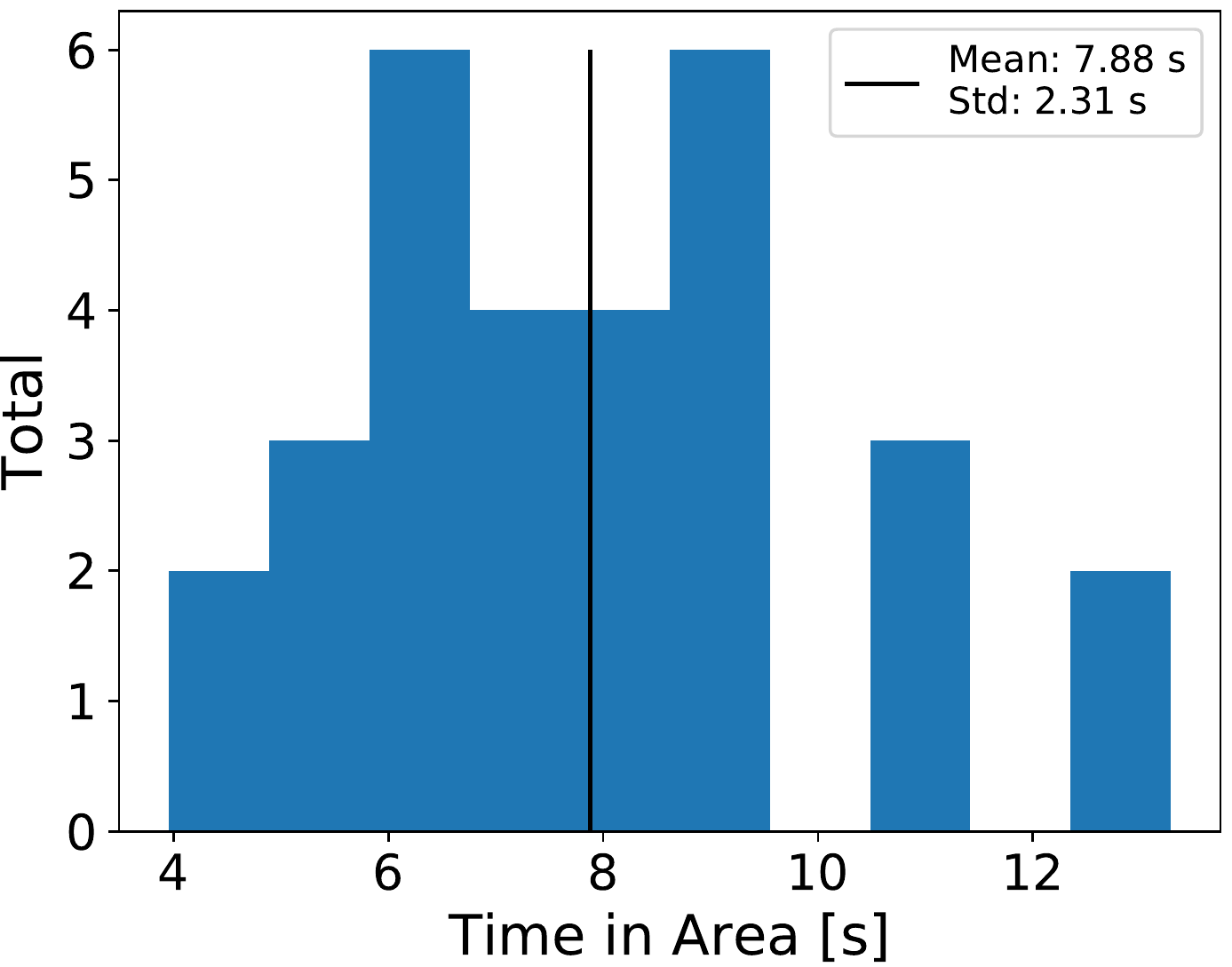}
    \caption{The duration of the walking participants inside the waiting area as histogram.}
    \label{fig:WaitingAreaTimeRunnerHistogram}
\end{figure}

Then we measured the time the walking participants spent in the rectangular waiting area of \(\SI{1.55}{\meter} \times \SI{1.7}{\meter}\) (width \(\times\) height), see \fref{fig:WaitingAreaTimeRunnerHistogram}.

The trajectory plots show that \textbf{all} walking participants were able to cross the waiting crowd. Instead of straight lines, we observe curvy trajectories where walking participants move around a waiting person or both seem to swap places.
\gertaaccepted{Our measurements of the waiting participants' maximum displacement in \fref{fig:ExperimentDenseCrowdWaitingCrowdMaxDistanceMetricBestDistributions},
and \fref{fig:SingleTrajectoryTimeStep0_8} show that the waiting participants also move. We argue, that this indicates interaction.} In fact, during the experiment we saw different techniques: communication through eye contact or asking verbally, but also shoving the waiting person aside. \gertaaccepted{Recent virtual reality experiments that track eye-gaze in dense crowds underline pedestrians' focus on the closest vicinity for interactions \cite{berton-2020}. This supports our hypothesis that collaboration with the next neighbors enables pedestrians to navigate through a dense crowd. We will use this finding to choose a suitable action in our model of cooperative behaviour.}

\fref{fig:WaitingAreaTimeRunnerHistogram} visualizes the duration
a walking participant spends inside the waiting area.
It indicates that the interaction process between the participants takes time. The mean duration of a walking participant's stay inside the waiting area is \SI{7.88}{\second}. Note, that if a walking participant walked through the waiting area, on a straight line, with an instantaneous speed of \SI{0.70}{\meter\per\second} (measurement from \tref{tbl:ExperimentDenseCrowdSpeedStatistic}), it would only take \( \frac{\text{height}}{\text{speed}} = \frac{\SI{1.70}{\meter}}{\SI{0.70}{\meter\per\second}} = \SI{2.43}{\second} \).

\section{Model}
\label{sec:Model}

\subsection{The Need for a Psychology Model Complementing Pure Locomotion}
\label{sec:ModelingOfPureLocomotion}

\begin{figure}[!h]
    \begin{subfigure}{\linewidth}
        \includegraphics[width=0.90\linewidth]{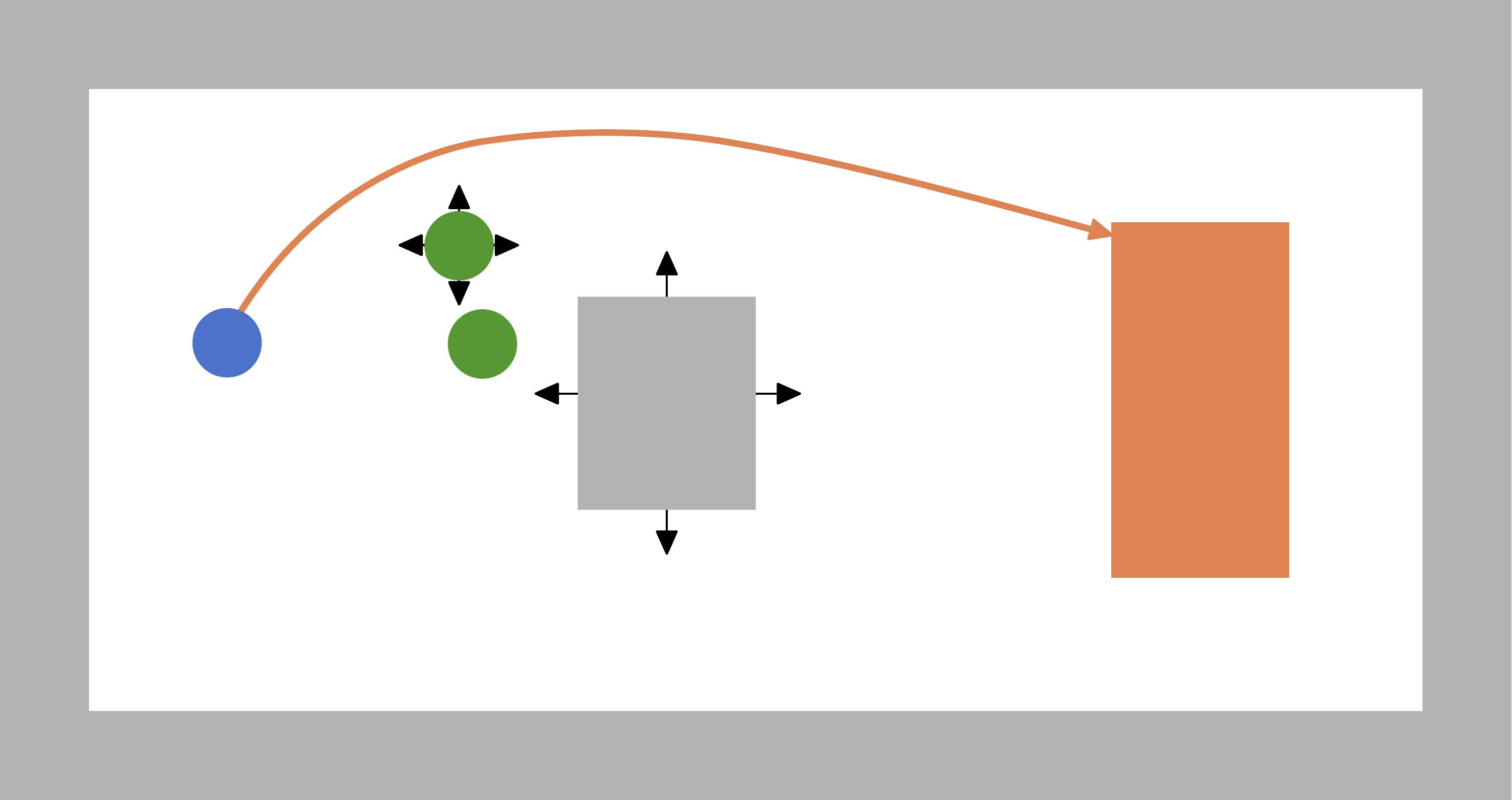}
        \caption{Modeling only locomotion: A walking agent (blue) is repelled (black arrows) by other agents (green) and obstacles (grey) and attracted by its target (orange). The repulsion is visualized by arrows.}
    \end{subfigure}\\
    \begin{subfigure}{\linewidth}
        \includegraphics[width=0.90\linewidth]{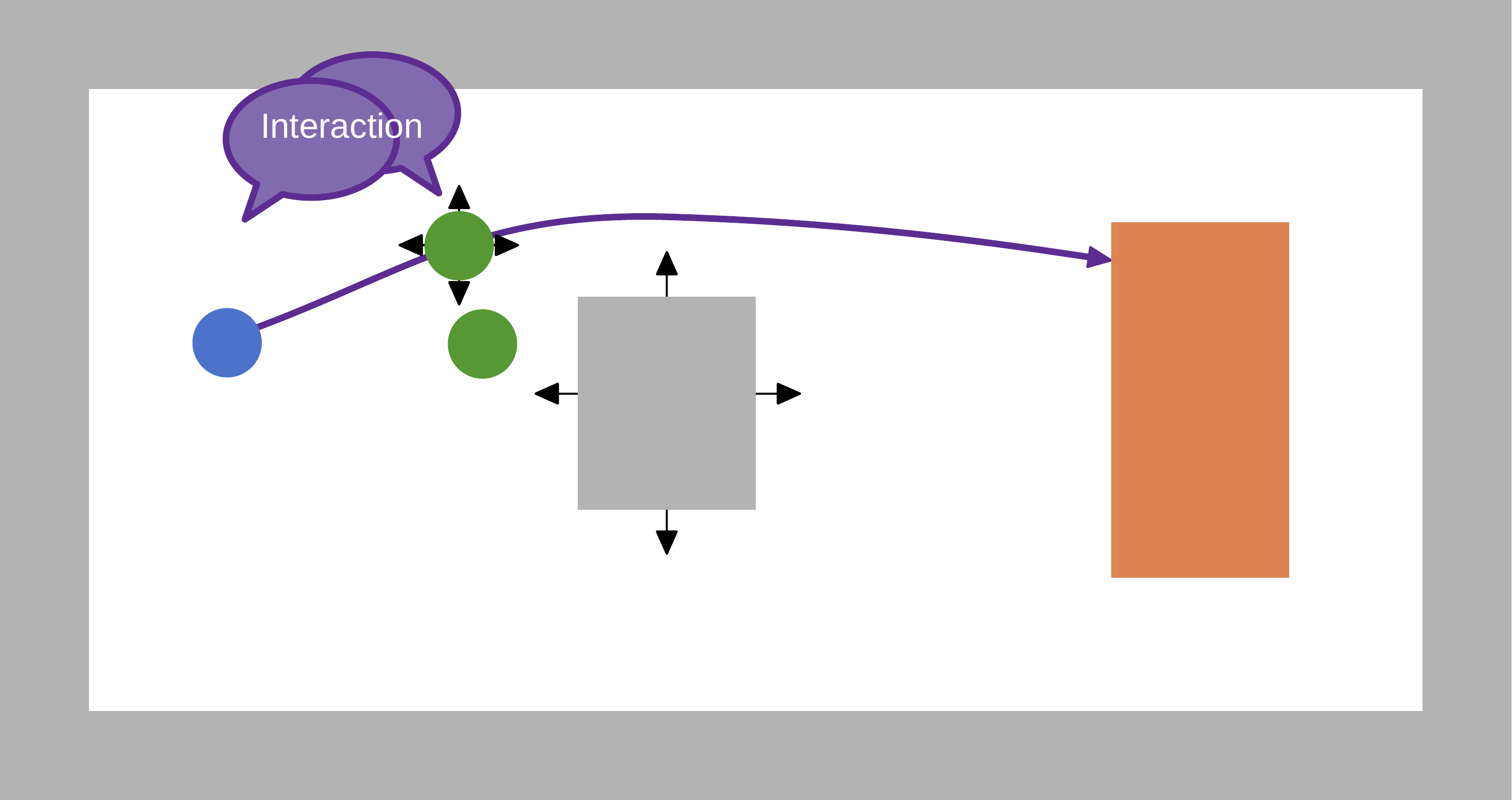}
        \caption{Modeling locomotion and psychology: A psychology layer allows agents to interact with each other.}
    \end{subfigure}
    \caption{Current pedestrian stream models focus on a pure locomotion of agents and neglect scientific findings from psychology like interaction between people and collective actions in crowds exemplified by \cite{templeton-2015}.}
    \label{fig:ModelingLocomotionAndPsychology}
\end{figure}

In the experiment, \textbf{all} walking participants were able to cross the waiting crowd by interacting with the other participants. From this, we derived our hypothesis that real humans can pass a crowd at high densities. \gertaaccepted{
This simple hypothesis is essential since it is where pedestrian stream simulators often fail, see \fref{fig:SimulationsAndBlockedAgent}. We will focus on this challenge with our new model proposal. Attempts to solve the problem solely on the locomotion layer, through collision avoidance as depicted in \fref{fig:ModelingLocomotionAndPsychology} do not work for very dense crowds which is shown by the simulations in 
\fref{fig:SimulationsAndBlockedAgent}}.
In the real world, however, humans adapt their behaviour \cite[p. 11--12]{gerrig-2013} to the situation. Humans use perception, cognition and a repertoire of different behaviours.
They also interact.

\gertaaccepted{We strive for a model that fulfils two important requirements:} \ii{1} Firstly, the new model shall represent a generic architecture which can be easily integrated into different simulation tools, independent of the choice of locomotion model, and that can be generalized to other instances of collective cooperation. Thus, the new model will be beneficial for the whole research community.\gertaaccepted{\ii{2} Secondly, the new model shall be a faithful operationalisation of psychological processes} which affect the behaviour of agents. That is, it must be correct from a psychological perspective and it must be sufficiently simple to be understood by different research communities like computer scientists, physicists, sociologists or psychologists.

\subsection{Model of a Psychology Layer for Collective Cooperation}
\label{sec:ModelPsychologyLayer}



Like for any other simulation software, a pedestrian stream simulator's core is a simulation loop in which time is incremented. In this loop, a locomotion model is responsible for finding the next position for each agent in each simulated time step (compare \lref{lst:SimulationLoop}).

\begin{lstlisting}[caption={A typical simulation loop of a pedestrian stream simulator.},label={lst:SimulationLoop},language=Java]
while (simulationIsRunning) {
  ...
  // A locomotion model searches the next
  // position for an agent which is closer
  // to a target than currently.
  locomotionModel.update(agents, time);
  ...
  time++;
}
\end{lstlisting}

Most of the current locomotion models \cite{helbing-1995,antonini-2006,seitz-2012,dietrich-2014} only include physical aspects to navigate an agent through an environment. For instance, obstacles repel an agent while targets attract agents.

But, the key is to include also the psychological status of an agent in each simulation step. This layer represents the mental processes of perception and cognition of real humans \cite[p. 206ff.]{gerrig-2013} and effects the behaviour of an agent. Additionally that means, instead of having just one behaviour --- \ie, moving towards a target --- an agent must have a behavioural repertoire from which the agent can choose from to react to its environment. In the case of our experiment, that means that agents (both walking and waiting)

\begin{itemize}
    \item on \textcolor{PerceptionColor}{perception sub-layer}, perceive other agents in a sight / search radius \(r\).
    \item on \textcolor{CognitionColor}{cognition sub-layer}, realise that an agent cannot move anymore (\ie, the speed over the last \(n\) steps is below a certain threshold) and they change their self-category \cite{reicher-2010} from target-oriented to cooperative to follow new social norms. We chose the term self-category here because \cite[p. 20]{templeton-2020} states that \enquote{self-categorisation [...] becomes the psychological basis for crowd behaviour} which is in our case collective cooperation.
    \item on \textcolor{LocomotionColor}{locomotion layer}, being cooperative means that agents swap places to reach their target.
\end{itemize}

\begin{figure}[!h]
    \centering
    \includegraphics[width=0.90\linewidth]{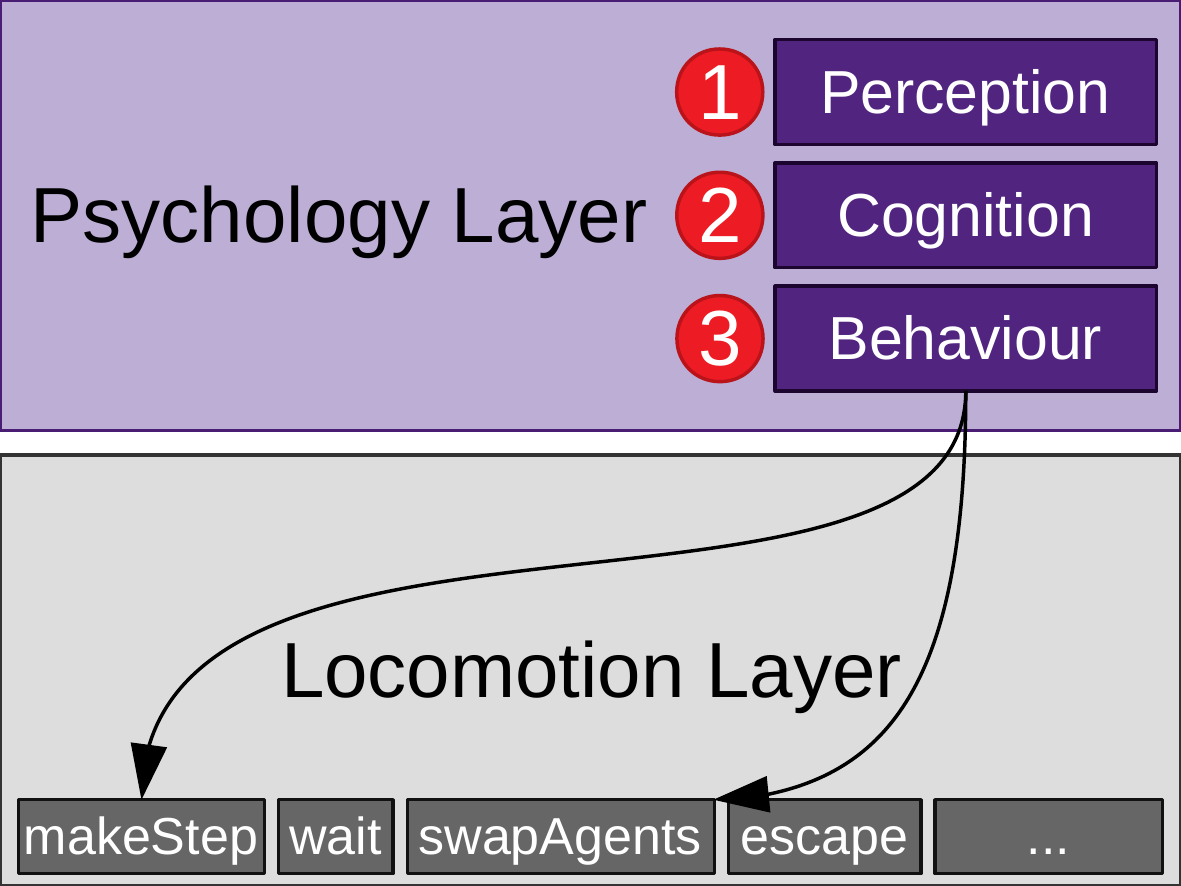}
    \caption{The three sequential phases of the new psychology layer: Firstly, agents perceive environmental stimuli. Secondly, agents process these information in a cognition phase and enrich it with further (context-relevant) information. Thirdly, agents react to the processed information by selecting a behaviour from a behavioural repertoire on locomotion layer. The behaviour repertoire on locomotion layer should cover different real-world situations. For instance, make a step towards a target (\eg, a train station), wait at a platform (\ie, do not move) or escape from a bang stimulus (which consists of several locomotion patterns).}
    \label{fig:VaderePsychologyLayerDetailedSteps}
\end{figure}

\fref{fig:VaderePsychologyLayerDetailedSteps} visualizes the sequential processing of information inside the introduced psychology layer. The lower layers, e.g. \code{Cognition}, process only the information from the direct upper layer. That means, an agent firstly perceives environmental stimuli, then an agent processes this information in the cognition layer and enriches it with further information (in case of the experiment this would be an agent's speed). This simple architecture reflects what real humans do: Perceive, process and react to this information with a specific behaviour. Look up figures in \sref{sec:ApplyingTheNewModel} to see the model in action and how target-oriented agents get cooperative and swap places to reach their target.

The main advantage of this clearly separated psychology layers is that experts in psychology or other fields can implement the perception and cognition sub-layer without knowing implementation details of the pedestrian stream simulator. A locomotion expert can implement the specific locomotion strategies. For instance, if cooperative behaviour does not mean swapping two agents, another locomotion strategy can be implemented on locomotion layer. This clean software architecture makes it possible to work interdisciplinary on a pedestrian stream simulator combining knowledge from different research domains like proposed by \cite[p. 46]{templeton-2020}.

Introducing this psychology layer (with sub-layers perception, cognition and locomotion) modifies the existing simulation loop \lref{lst:SimulationLoop} only very slightly and keeps the overall software architecture simple and easy to implement according to the KISS principle \cite[p. 18]{axelrod-1997} \cite[p. 10]{martin-2008}, compare \lref{lst:SimulationLoopWithPsychologyLayer}.

\begin{lstlisting}[caption={The new simulation loop which contains the added psychology layer with sub-layers perception, cognition and behaviour.},label={lst:SimulationLoopWithPsychologyLayer},language=Java]
while (simulationIsRunning) {
  ...
  %{\color{PerceptionColor}
  // Perception
  }%
  perceptionModel.update(agents, stimuli);
  ...
  %{\color{CognitionColor}
  // Cognition
  }%  
  cognitionModel.update(agents);
  ...
  %{\color{LocomotionColor}
  // Locomotion
  }%
  locomotionModel.update(agents, time);
   |
   +-> if (agent.selfCategory == COOPERATIVE) {
         Agent candidate = findSwapCandidate();
         swapAgents(agent, candidate);
       }
  ...
  time++;
}
\end{lstlisting}

\begin{figure}[!h]
    \begin{subfigure}{\linewidth}
        \includegraphics[width=0.78\linewidth]
        {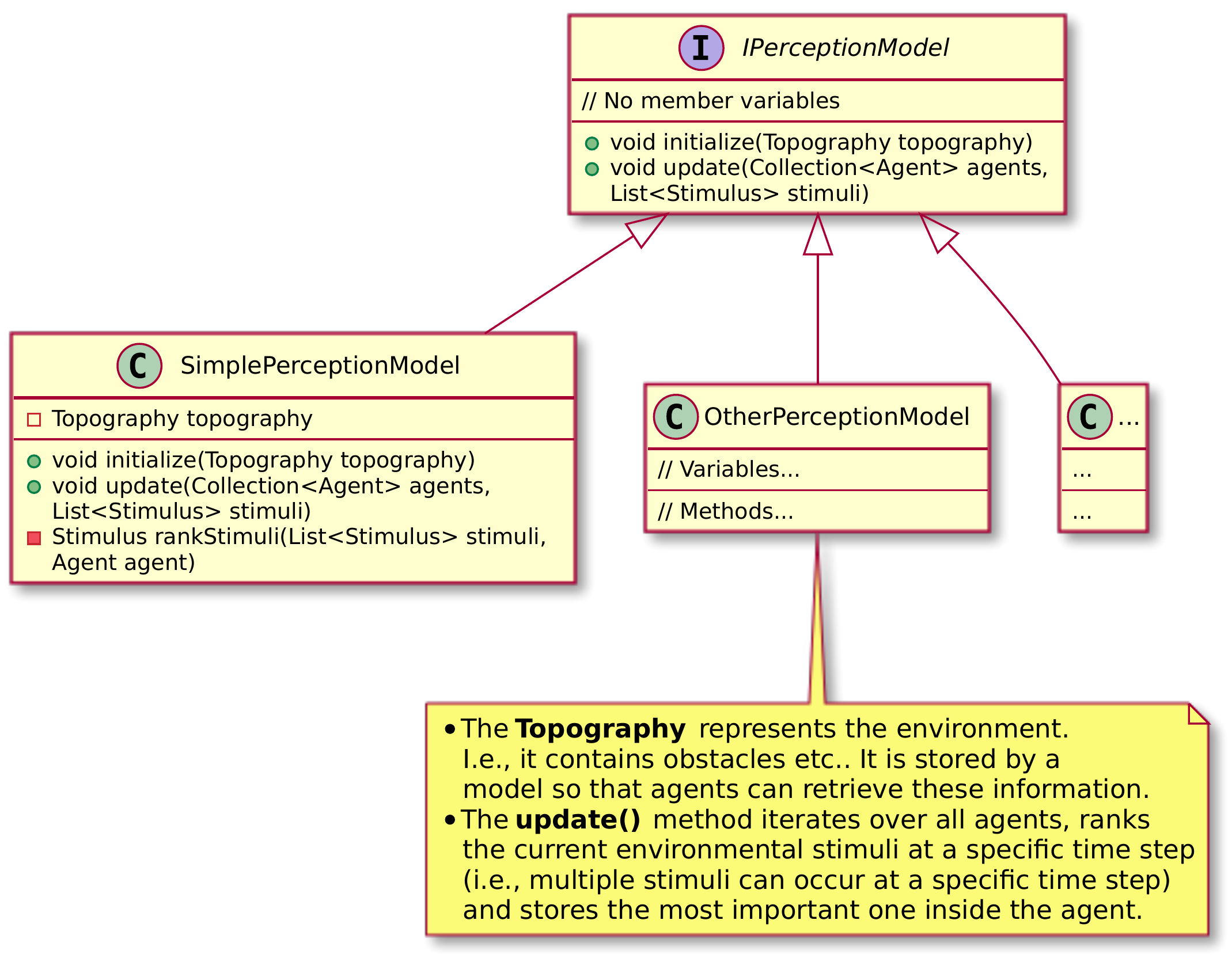}
        \caption{Interfaces and classes of the perception sub-layer.}
    \end{subfigure}\\
    \begin{subfigure}{\linewidth}
        \includegraphics[width=0.78\linewidth]{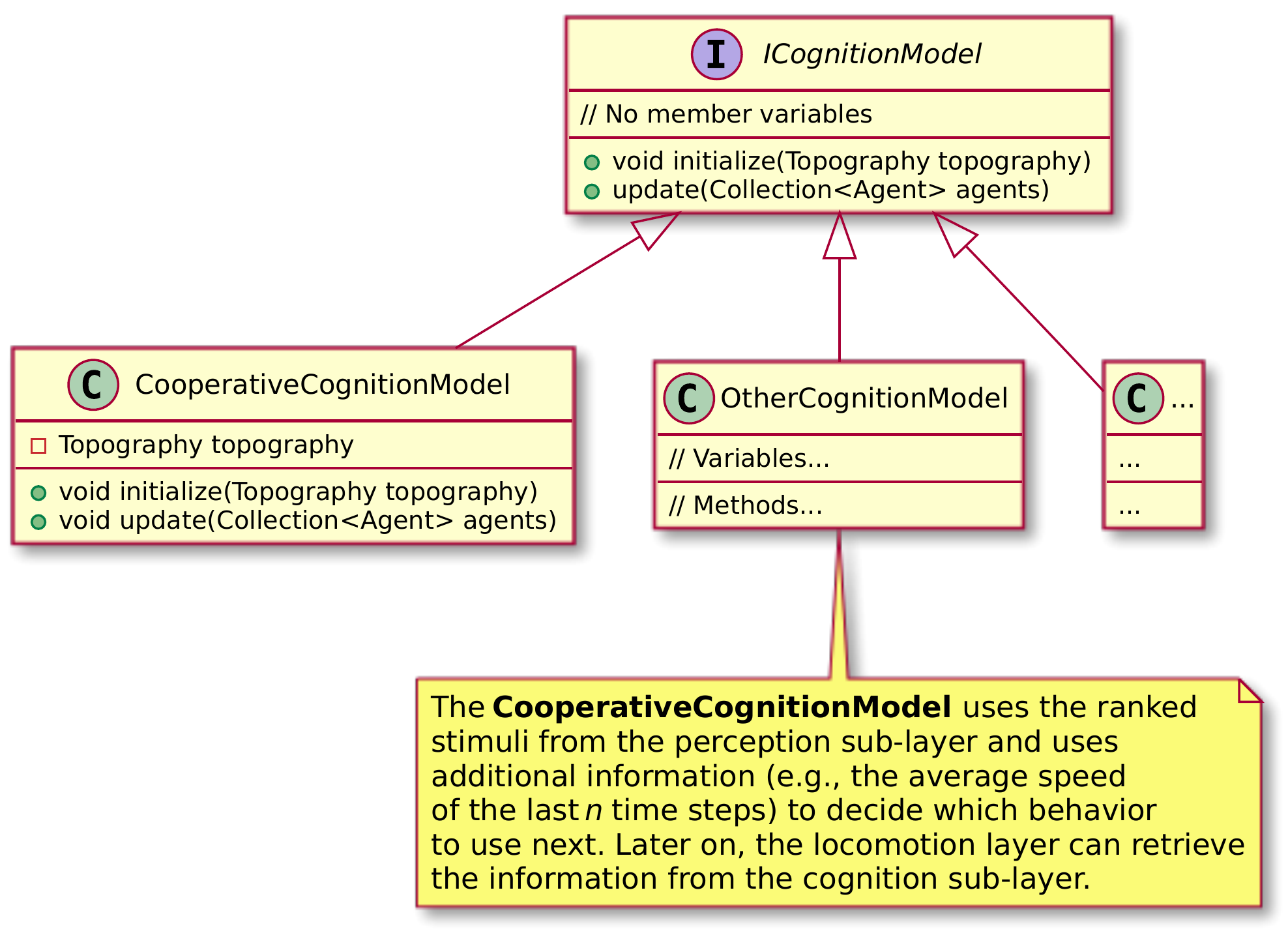}
        \caption{Interfaces and classes of the cognition sub-layer.}
    \end{subfigure}\\
    \caption{An UML diagram of the interfaces and classes of the perception and cognition sub-layer. Public methods are denoted with green circles, private methods are denoted with filled, red squares and member variables are denoted with unfilled, red squares.}
    \label{fig:UMLDiagramsPerceptionAndCognition}
\end{figure}

\code{perceptionModel} and \code{cognitionModel} are implementations of interfaces. Using this design decision --- the strategy pattern --- allows to extend a pedestrian stream simulator to a tool to test also psychological hypothesis. That means that it is possible to change the perception and cognition model for each simulation run and allows to cover different real-world situations. For instance, an experiment situation differs from a daily commuting situation which affects humans' perception and cognition. This reflects also the fact that a simulation tool cannot provide a \enquote{one-fits-all-situations} model. Therefore, we facilitate interfaces with only two methods here, see UML diagram in \fref{fig:UMLDiagramsPerceptionAndCognition}.

\lref{lst:CognitionCooperativeCognitionModel} and \lref{lst:LocomotionModel} shows that it only requires 13 lines on cognition sub-layer and 24 lines on locomotion layer to get collective cooperative agents and to reenact the experiment. E.g., if an agent (walking and waiting) cannot move anymore, it gets cooperative. Cooperative behaviour results in swapping positions.

\begin{lstlisting}[caption={The \code{update()} method of class \code{CooperativeCognitionModel} which toggles an agent's self category from target-orientied to cooperative based on agent's speed to reenact the experiment.},label={lst:CognitionCooperativeCognitionModel},language=Java]
int lastSteps = 4;
double threshold = 0.05;

for (Agent agent : agents) {
  boolean cannotMove =
    agent.getSpeed(lastSteps) <= threshold

  if (cannotMove) {
      agent.setSelfCategory(COOPERATIVE);
  } else {
      agent.setSelfCategory(TARGET_ORIENTED);
  }
}
    
\end{lstlisting}

\begin{lstlisting}[caption={The \code{update()} and \code{updateAgent()} method of the locomotion model which reacts to agent's psychology status reflected by \code{agent.getSelfCategory()}.},label={lst:LocomotionModel},language=Java]
public void update(Agents agents, double time) {
  for (Agent agent : agents) {
      updateAgent(agent, time)
  }
}

void updateAgent(Agent agent, double time) {
  selfCategory = agent.getSelfCategory();
  
  if (selfCategory == TARGET_ORIENTED) {
      makeStepToTarget(agent);
  } else if (selfCategory == COOPERATIVE) {
      // Search for other cooperative agents
      // in a search radius r.
      Agent candidate = findSwapCandidate(agent);
  
      if (candidate != null) {
          swapPedestrians(agent, candidate);
      } else {
          makeStepToTarget(agent);
      }
  }
  ...
}
\end{lstlisting}

The proposed psychology layer was implemented in \program{Vadere} \cite{kleinmeier-2019,vadere-2020} because it is open source and has already a well-validated locomotion layer \cite{seitz-2016,seitz-2016,dietrich-2017}. Vadere is an open source framework for the simulation of microscopic pedestrian dynamics.

The following steps were carried out:

\begin{enumerate}
    \item Add interfaces \code{IPerceptionModel} and \code{ICognitionModel} (see UML diagrams in \fref{fig:UMLDiagramsPerceptionAndCognition}).
    \item To reenact the experiment setup from \sref{sec:PresentationSetUp}, implement \code{SimplePerceptionModel} and \code{CooperativeCognitionModel}. \code{SimplePerceptionModel} is empty because there were no external stimuli present in the experiment. \code{CooperativeCognitionModel} changes an agent's self category from target-oriented to cooperative if an agent cannot move anymore (\ie, its speed is below a certain threshold by storing an agent's psychology status with \code{agent.setSelfCategory(SelfCategory newSelfCategory)}, see \lref{lst:CognitionCooperativeCognitionModel}.
    \item Extend the existing simulation loop: In each simulation loop, invoke \code{perceptionModel.update()} and \code{cognitionModel.update()}, see \lref{lst:SimulationLoopWithPsychologyLayer}.
    \item On locomotion layer, evaluate \code{agent.getSelfCategory()} and react to it, see \lref{lst:LocomotionModel}.
\end{enumerate}

\subsection{Re-enacting the Experiment with the Collective Cooperation Model}
\label{sec:ApplyingTheNewModel}

\begin{figure}[!h]
    \begin{subfigure}{0.20\linewidth}
        \includegraphics[width=\linewidth]{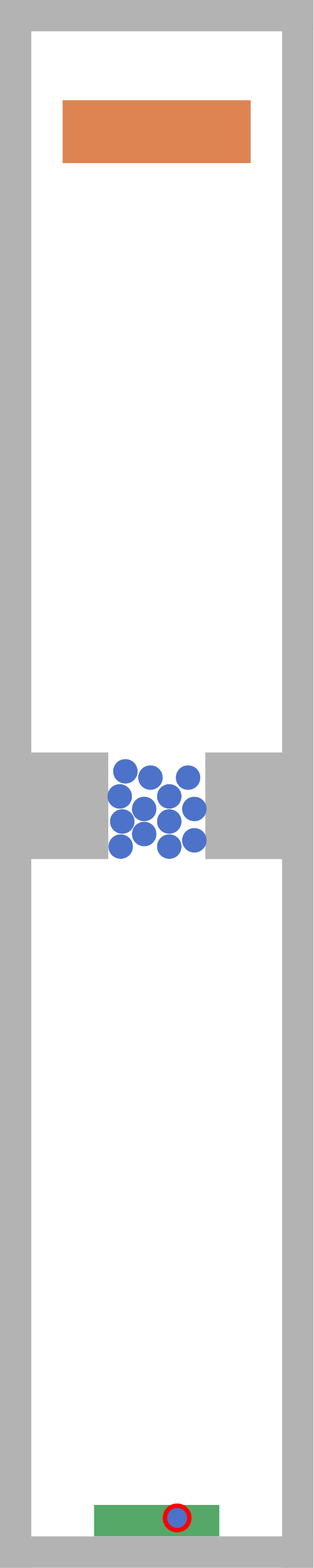}
        \caption{Time step 1}
    \end{subfigure}
    \begin{subfigure}{0.20\linewidth}
        \includegraphics[width=\linewidth]{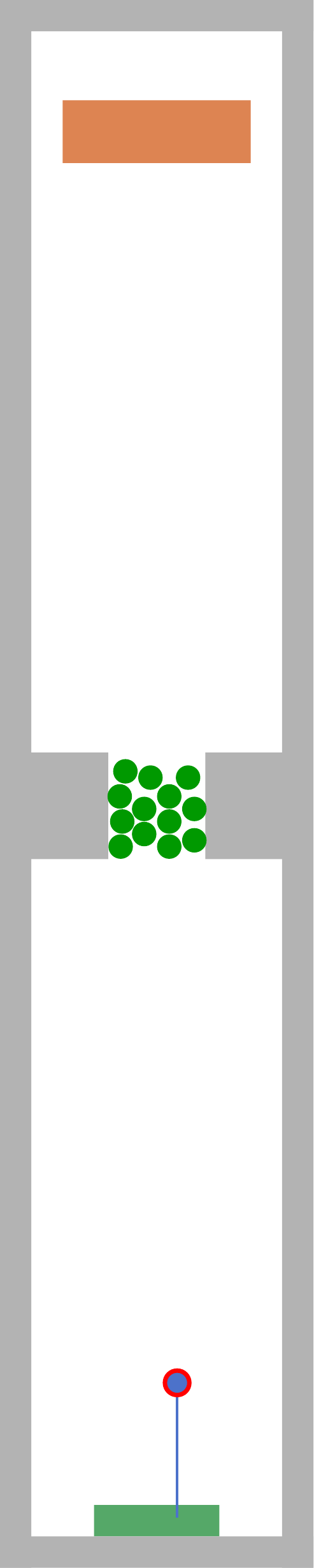}
        \caption{Time step 4}
    \end{subfigure}
    \begin{subfigure}{0.20\linewidth}
        \includegraphics[width=\linewidth]{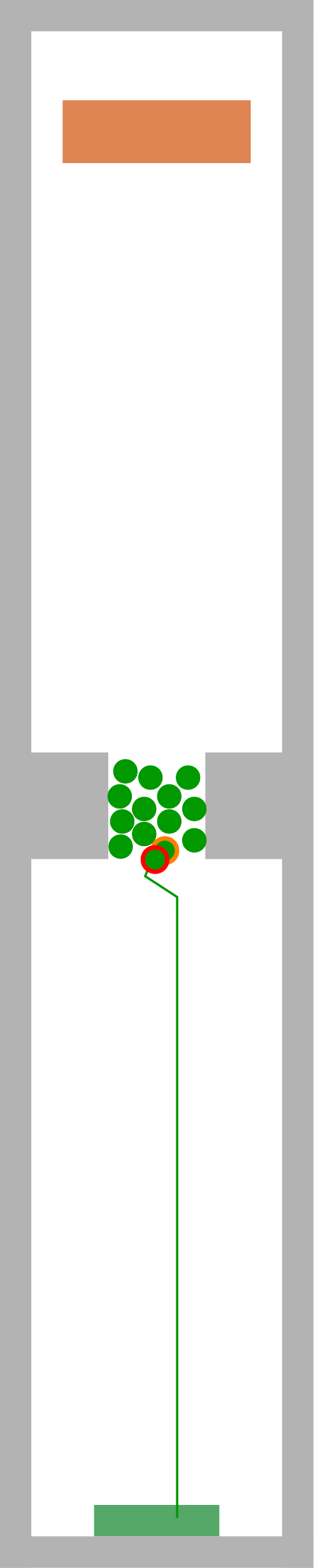}
        \caption{Time step 29}
    \end{subfigure}
    \caption{A walking agent (red-encircled) starts walking in the green source area and tries to reach the brown target area while the agent is blocked by a waiting crowd consisting of 13 agents. The colours represent the current behaviour of an agent: \textcolor{TargetOrientedColor}{Blue} is \textcolor{TargetOrientedColor}{target-oriented behaviour} and \textcolor{CooperativeColor}{green} is \textcolor{CooperativeColor}{cooperative behaviour}. \ii{Time step 1} When the simulation starts, all agents are target-oriented. While the walking agent is attracted by the brown target, the waiting crowd does not have a target and waits. \ii{Time step 4} The agents of the waiting crowd get cooperative because their speed falls below a certain threshold. \ii{Time step 29} The walking agents reaches the waiting crowd and cannot move anymore. Thus, the walking agent also gets cooperative. The walking agent searches for a swap candidate (orange-encircled) and both swap positions.}
    \label{fig:VadereExperimentSetup1}
\end{figure}

\begin{figure}[!h]
    \begin{subfigure}{0.20\linewidth}
    \includegraphics[width=\linewidth]{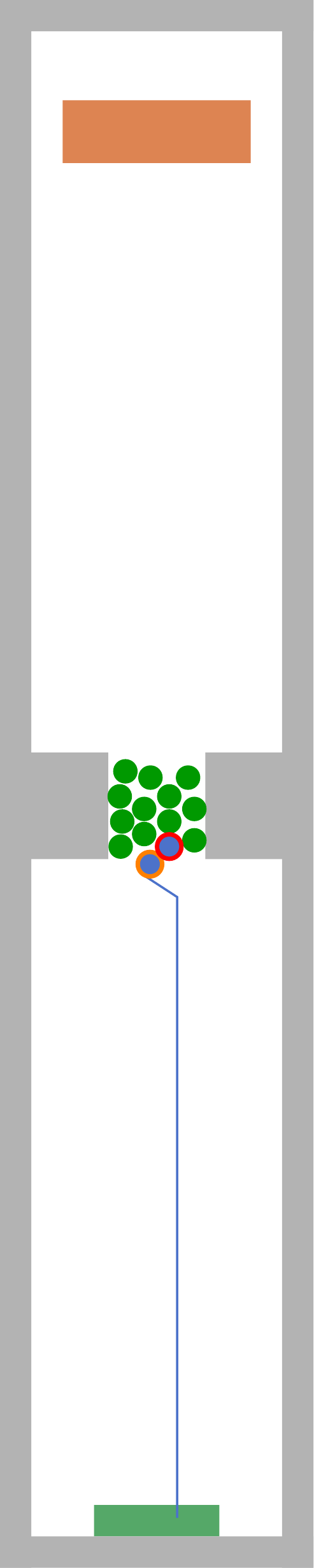}
    \caption{Time step 31}
    \end{subfigure}
    \begin{subfigure}{0.20\linewidth}
        \includegraphics[width=\linewidth]{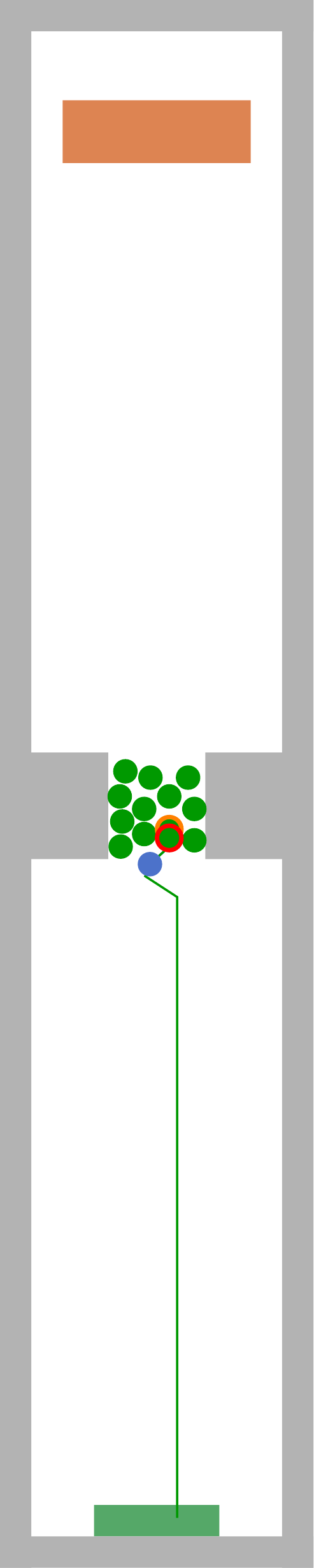}
        \caption{Time step 36}
    \end{subfigure}
    \begin{subfigure}{0.20\linewidth}
        \includegraphics[width=\linewidth]{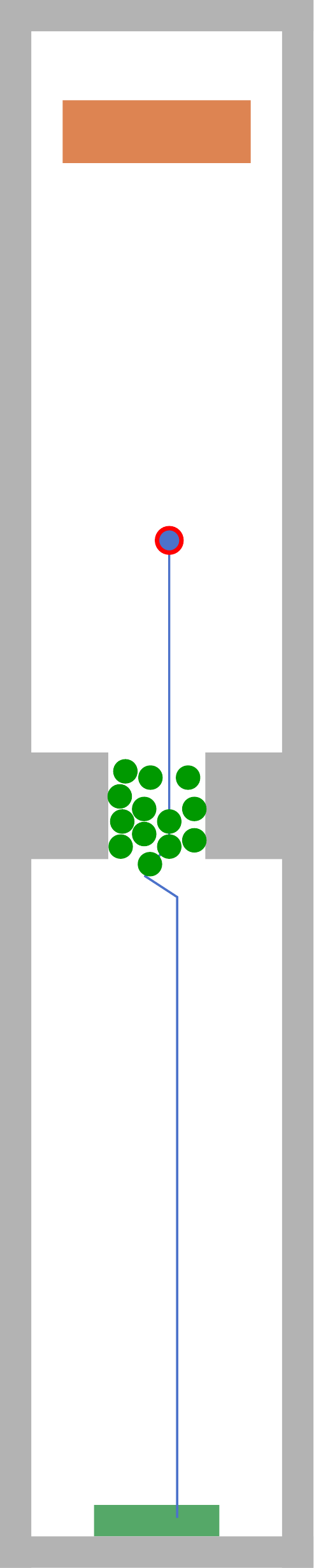}
        \caption{Time step 51}
    \end{subfigure}
    \caption{Cooperative behaviour of agents inside the waiting crowd. The colours represent the current behaviour of an agent: \textcolor{TargetOrientedColor}{Blue} is \textcolor{TargetOrientedColor}{target-oriented behaviour} and \textcolor{CooperativeColor}{green} is \textcolor{CooperativeColor}{cooperative behaviour}. \ii{Time step 31} After swapping positions, the walking agent (red-encircled) and the swap candidate (orange-encircled ) get target-oriented again because their speed is above a certain threshold. \ii{Time step 36} The walking agents gets cooperative again and swaps position with another cooperative agent which is closer to the target. \ii{Time step 51} The walking agent found its way through the dense crowd by using a cooperative behaviour.}
    \label{fig:VadereExperimentSetup2}
\end{figure}

The psychology layer was implemented in the pedestrian stream simulator \program{Vadere}. We reenacted the experiment setup from \sref{sec:PresentationSetUp} as closely as possible by using the same dimensioning. We carried out 100 simulation runs with slightly varying initial position of the walking agent but consistent positions for the agents of the waiting crowd. \fref{fig:VadereExperimentSetup1} and \fref{fig:VadereExperimentSetup2} show one of these simulation runs and visualize how the walking agent (red-encircled) changes its target-oriented behaviour to a cooperative one when the agent is blocked by the waiting crowd.


To validate the simulations, we compare the simulation results to the experiment results. In \sref{sec:ExperimentResults}, we measured the speed of the walking participant, the spatial distribution of the waiting crowd and the trajectories of walking participant. In our comparison, we omit the spatial distribution of the crowd because, in the implemented model the agents of the waiting crowd just wait in the waiting area and do not move at all. This is what we assumed as --- very simplified --- waiting behaviour. \Ie, the traveled distance of the agents of the waiting crowd is zero. Therefore, it makes no sense to compare it with the experiment participants which moved continuously at least a bit.

\begin{figure}[!h]
    \centering
    \includegraphics[width=\linewidth]{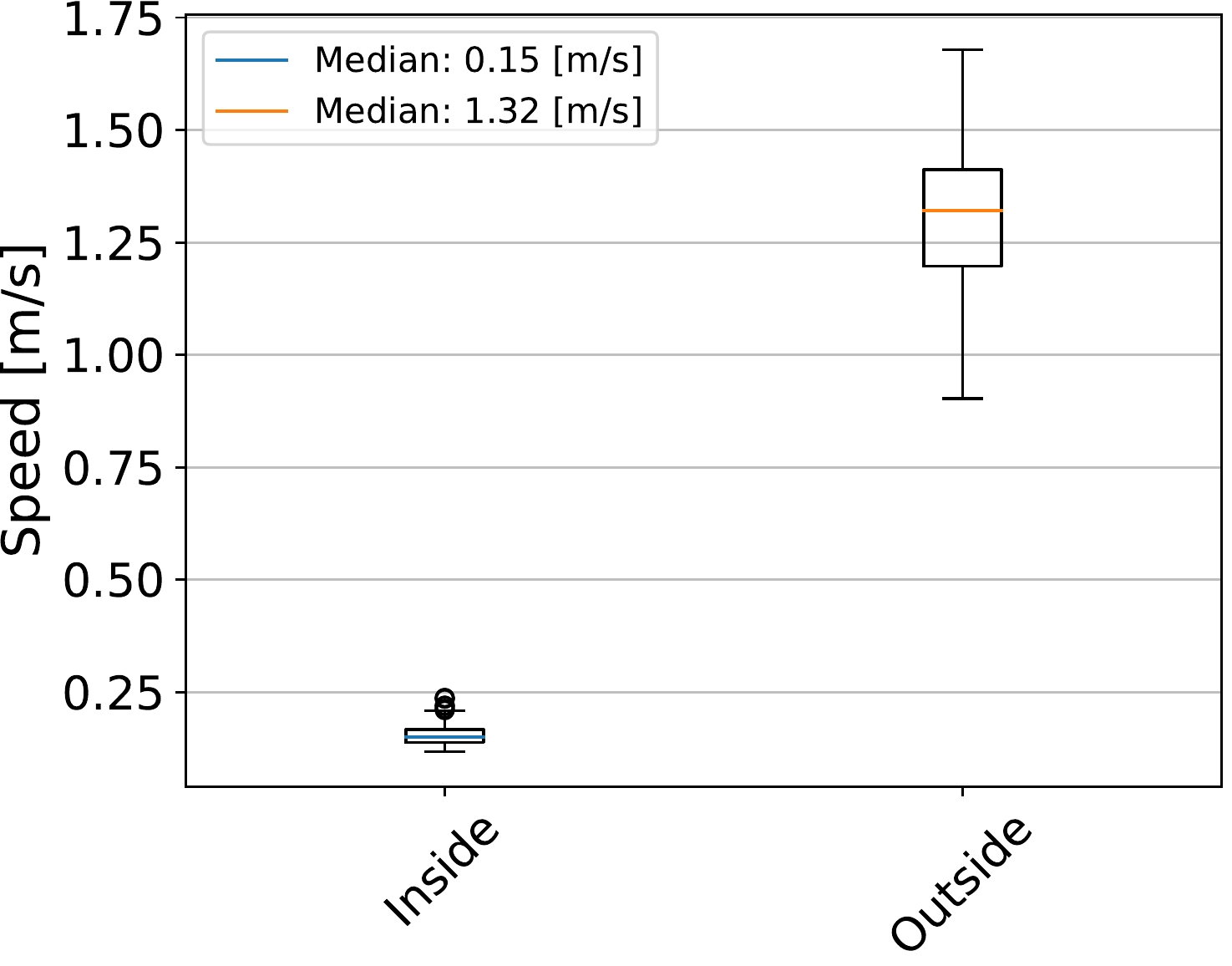}
    \caption{Box plot for speed distribution of the walking agent inside and outside the waiting crowd}
    \label{fig:SimulationDenseCrowdSpeedBoxplot}
\end{figure}

The 100 simulations reproduce the measured instantaneous \enquote{free-flow} speeds at least qualitatively: The walking agents are slowed down inside the waiting area from \SI{1.31}{\meter\per\second} (outside) to \SI{0.16}{\meter\per\second} (inside) on average compared to \SI{1.33}{\meter\per\second} and \SI{0.70}{\meter\per\second} in the experiment, see \fref{fig:SimulationDenseCrowdSpeedBoxplot} and     \tref{tbl:SimulationDenseCrowdSpeedStatistic}.

\begin{table}[!h]
    \begin{tabular}{lrr}
        \toprule
        {} &         \multicolumn{2}{c}{\textbf{Speed [\si{\meter\per\second}]}} \\
        {} &         \textbf{Inside} &  \textbf{Outside} \\
        \midrule
        sample size &            100.00 &           100.00 \\
        mean  &              0.16 &             1.31 \\
        std   &              0.03 &             0.15 \\
        min   &              0.12 &             0.90 \\
        25\%   &              0.14 &             1.20 \\
        50\%   &              0.15 &             1.32 \\
        75\%   &              0.17 &             1.41 \\
        max   &              0.24 &             1.68 \\
        \bottomrule
    \end{tabular}
    \caption{Detailed statistics for the measured speed distributions of the walking participants inside and outside the waiting crowd.}
    \label{tbl:SimulationDenseCrowdSpeedStatistic}
\end{table}

The speed of the walking agent inside the waiting crowd is much lower than what we have observed in the experiment. In the experiment, even if the walking participant is blocked by the waiting crowd for some moments, the walking participant constantly moves its body a tiny bit. That means the speed of the walking participant is constantly greater than zero. But in the simulation, it takes some simulation steps until a walking agent gets cooperative when the agent is blocked by the waiting crowd. \Ie, the agent's speed is zero for a lot of simulation steps which lowers the average speed of the walking agents. Please keep in mind that this is the very first version of such a psychological model of collective cooperation and it will require some sort of calibration in the future.

\begin{figure}[!h]
    \centering
    \includegraphics[width=\linewidth]{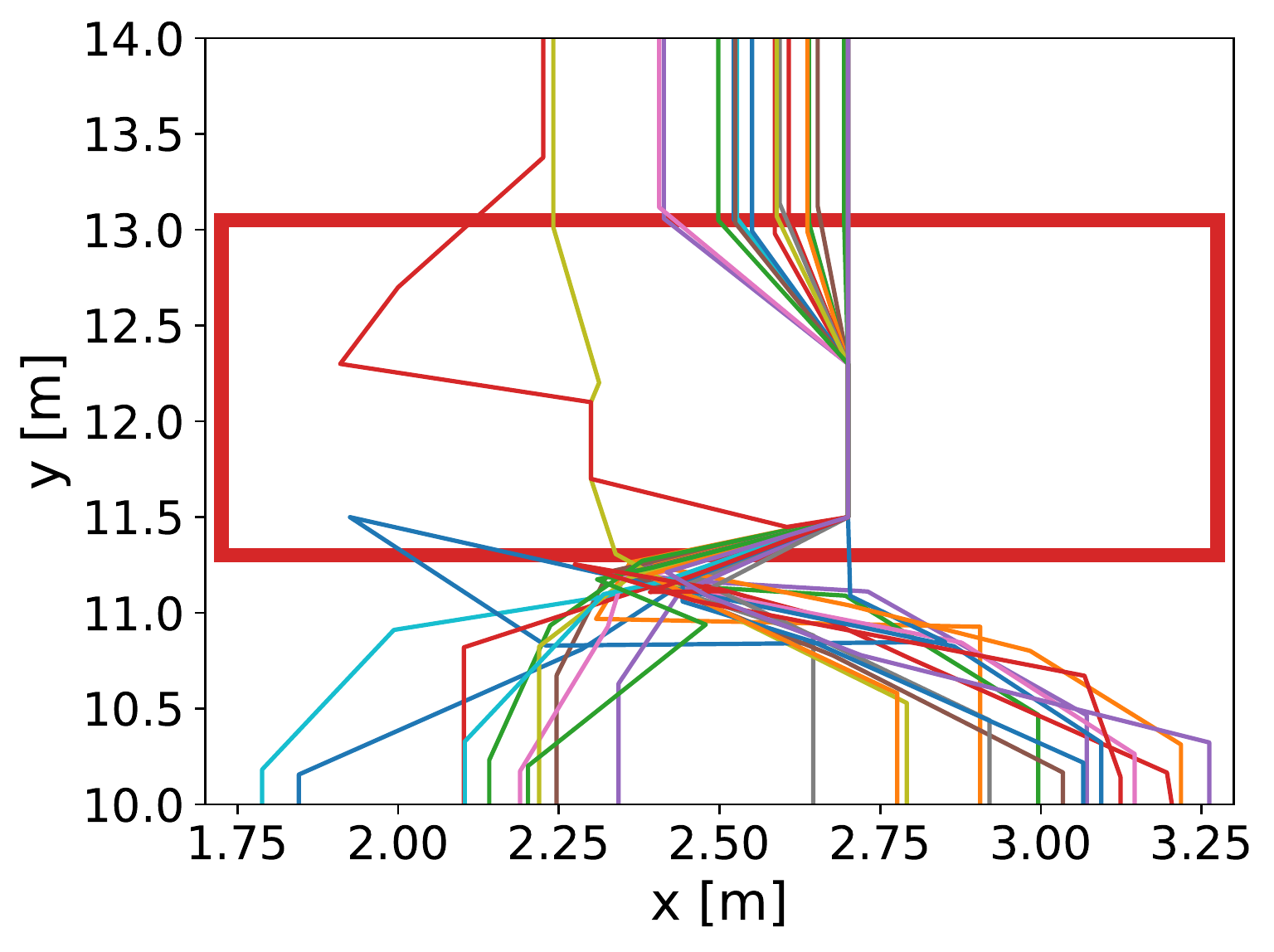}
    \caption{The trajectories of 25 walking agents inside the waiting area (red rectangle). Inside the waiting area, the walking agents follow zig-zag trajectories because they swap positions with agents of the waiting crowd. By changing to a cooperative behaviour, all walking agents were able to reach the target region. The agents of the waiting crowd are placed at the same positions for all 100 simulation runs. Therefore, we did not see a greater variety of the trajectories inside the waiting area.}
    \label{fig:SimulationTrajectoriesAllStepSize4}
\end{figure}

\begin{figure}[!h]
    \centering
    \includegraphics[width=\linewidth]{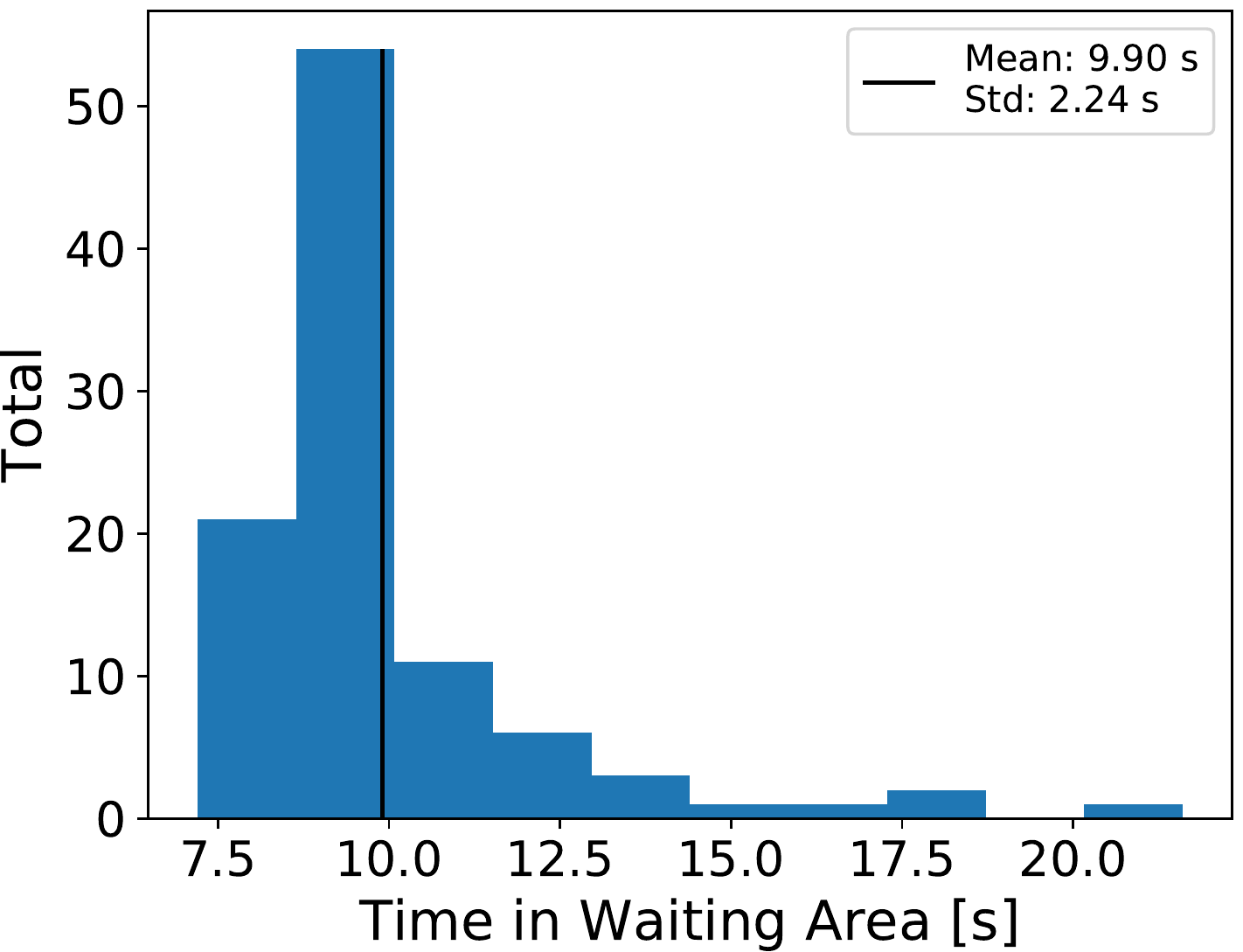}
    \caption{The duration of the walking agents inside the waiting area as histogram.}
    \label{fig:SimulationWaitingAreaTimeRunnerHistogram}
\end{figure}

Nevertheless, in our simulations we see that \textbf{all} walking agents were able to cross the waiting crowd like in the experiment with real humans, see \fref{fig:SimulationTrajectoriesAllStepSize4}. Also the mean time of the walking agent inside the waiting area is very close to the experiment observations: \SI{9.90 +- 2.24}{\second} in simulation compared to \SI{7.88 +- 2.31}{\second} in the experiment, see \fref{fig:SimulationWaitingAreaTimeRunnerHistogram}.

\section*{Conclusion}
\label{sec:Conclusion}

\gertaaccepted{
We identified a major shortcoming in current pedestrian simulation models: The lack of collective cooperation which means that agents fail at seemingly simple tasks, such as forging a path through a dense crowd. Since empirical evidence is extremely scarce we
also presented a controlled experiment to observe what really happens when participants pass a waiting crowd: We placed students in a delimited area of \SI{2.64}{\meter\squared} and let other participants walk through  this waiting crowd. We took measures to avoid observer biases and to obtain reliable data from the experiment.} We derived three hypotheses, namely: 
\ii{1} Real humans succeed in passing through a crowd at high density.
\ii{2} Pedestrians walking through a crowd are slowed down.
\ii{3} The pedestrians in the waiting crowd mostly return to their initial positions after giving way to the individual they allow through. While seemingly trivial, the first hypothesis is vital, because this is where classic locomotion models fail.

We presented a model where agents interact with each other to allow collective cooperation. Agents are able to perceive their environment, process this information and enrich it with additional information (within the simulation) in a cognition process. Then, agents select from a portfolio of possible behaviours, notably being target-oriented, or being cooperative. Actions on the locomotion layer follow, such as making a step towards a target or swapping place with a cooperation partner. The model is independent of the choice of locomotion model. It's implementation within the Vadere simulation framework is free and open-source.

In a reenactment of the experiment situation, our simulations
qualitatively reproduce the empirical observations. Most importantly,
agents' ability to pass through a dense crowd emerges as an effect of the psychological model.  

In the future we hope that new experiments and field observations will
bring more qualitative and quantitative information on behaviours in dense  crowds so that we --- or other modellers --- may add to the portfolio of behaviours, and calibrate parameters for a better quantitative fit. Further, our generic approach allows to cover a wide range of real-world situations with collective behaviours. As a next step, we would like to mold people's collective reaction to perceived threats within the same framework.

 
\section*{Supplement}
\label{sec:Supplement}

\subsection{Experiment: Limitations}
\label{sec:ExperimentLimitations}

While running the two experiments we were faced with several problems, on which we would like to comment.
\iih{1}{Filming from above} Filming the scene with an angle relative to the vertical axis, as we had to do it in our setting, causes problems when tracking the people. They can disappear behind each other. Also there is more noise, than with filming the experiment from above.
\iih{2}{Automated trajectory extraction} The trajectory of the walking participants is extracted in a semi-automatic procedure using \program{Tracker's Auto-Tracking} feature. 
To facilitate automatic trajectory extraction \cite{boltes-2013,boltes-2016}, the walking participants could have worn coloured hats. Then, tracker software like \code{PeTrack} could be used to extract trajectories automatically. Additionally, it would be useful to print the participants id on the hat to be able to match the corresponding participant data like size and age. 
\iih{3}{Use more diverse participants} All participants were first semester students in their second study week. It would be useful to have more diverse participants to generalize findings to a broader population. 
\iih{4}{More data} 
The experiment yielded 30 runs in total. Therefore, it would be useful to have more experiment runs to get results which are statistically more significant. \iih{4}{Controlled experiment} We conducted a controlled experiment instead of a field experiment because it was easier to carry out in regard of legal concerns. The artificiality of the situation might have affected each participant's behaviour. Therefore, we would like to encourage the scientific community to replicate the experiment to support our findings or make new ones.

\subsection{Experiment: Trajectory extraction and error}
\label{sec:ExperimentMethods}

In this section, we describe how we extract participants' trajectories from the video footage and quantify the corresponding measurement error.

\subsubsection{Trajectory Extraction}
\label{sec:TrajectoryExtraction}

For the trajectory extraction we used mainly two software tools: \program{ffmpeg} and \program{Tracker}.

The open-source software \code{ffmpeg} was used to cut the videos into short sequences of around 15 seconds to show only a single experiment run. For this, we watched the video material manually and noted down the times when a walking participant entered and left the camera cutout. This list of start and end times was fed to a self-written Python script which invokes \code{ffmpeg} in turn.

The video analysis and modelling tool \program{Tracker} offers two possibilities to track objects in a video, either manually or automatically. With the manual approach, a user has to manually mark the position of the person of interest in each single video frame. The video material contains \SI{25}{frames\ per\ second}. Tracker's \enquote{Auto-Tracker} feature allows to track an object automatically. For this, the walking participant and the participants of the waiting crowd are marked as distinct features of interest in the video footage and then \enquote{Auto-Tracker} searches each following video frame for the best match to that template and stores the position and other information of the tracked object. Regardless of manual or automatic tracking, \program{Tracker} stores position, velocity and acceleration for each tracked object for each frame (\ie, every \SI{1/25}{\second}).

We applied the automatic trajectory extraction since it yielded a smaller measurement error (see next section)

\subsubsection{Trajectory Extraction: Measurement Error}
\label{sec:TrajectoryExtractionMeasurementError}

\begin{figure}[!h]
    \centering
    \includegraphics[width=\linewidth]{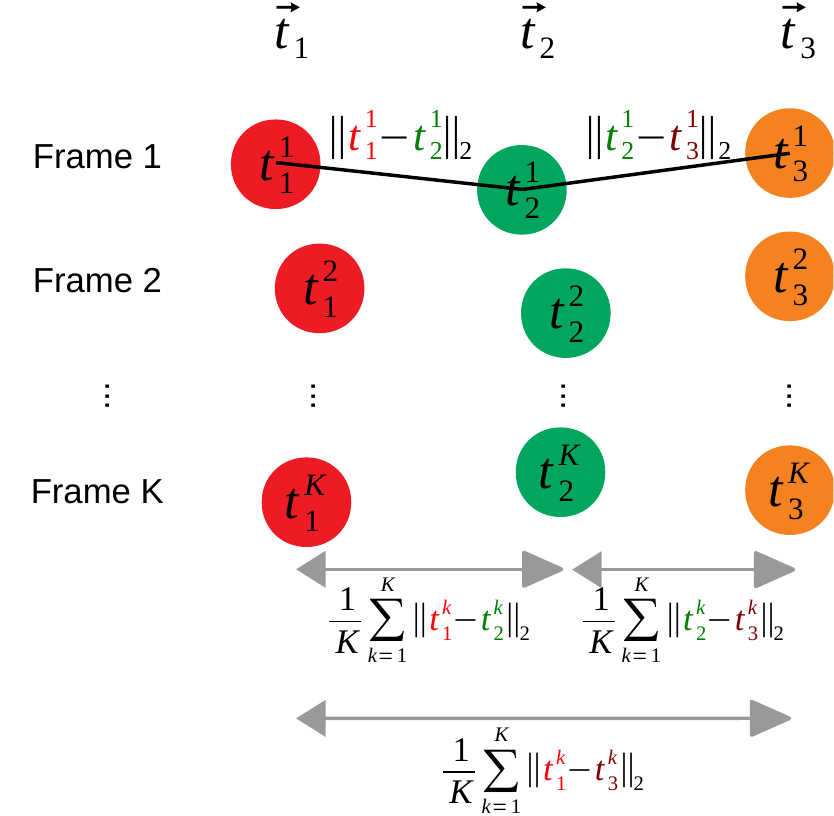}
    \caption{Simplified example to calculate the Euclidean distance between trajectory pairs. The actual measurement error is the mean Euclidean distance between all trajectory pairs.}
    \label{fig:MeasurementErrorDeltaTrajectories}
\end{figure}

To reveal the measurement error which is introduced by the trajectory extraction, four experiment assistants tracked the same person several times, both manually and automatically. Then, we cross-checked the automatically tracked trajectories against each other and in another comparison we cross-checked the manually tracked trajectories, see \fref{fig:MeasurementErrorDeltaTrajectories} as simplified example.

A trajectory \(i\) contains for each frame \(k\) the x and y coordinate as vector \(t_{i}^{k}\). As first step, we calculate the mean Euclidean distance for all \(K\) frames between two trajectories \(i\) and \(j\). As second step, we calculate the mean for all these trajectory pairs N:

\begin{equation}
\frac{1}{N} \sum_{i=1,j=i+1}^{N} \frac{1}{K} \sum_{k=1}^{K} \norm[2]{t_{i}^{k} - t_{j}^{k}}
\end{equation}

\begin{figure}[!h]
    \centering
    \begin{subfigure}{0.8\linewidth}
        \centering
        \includegraphics[width=\linewidth]{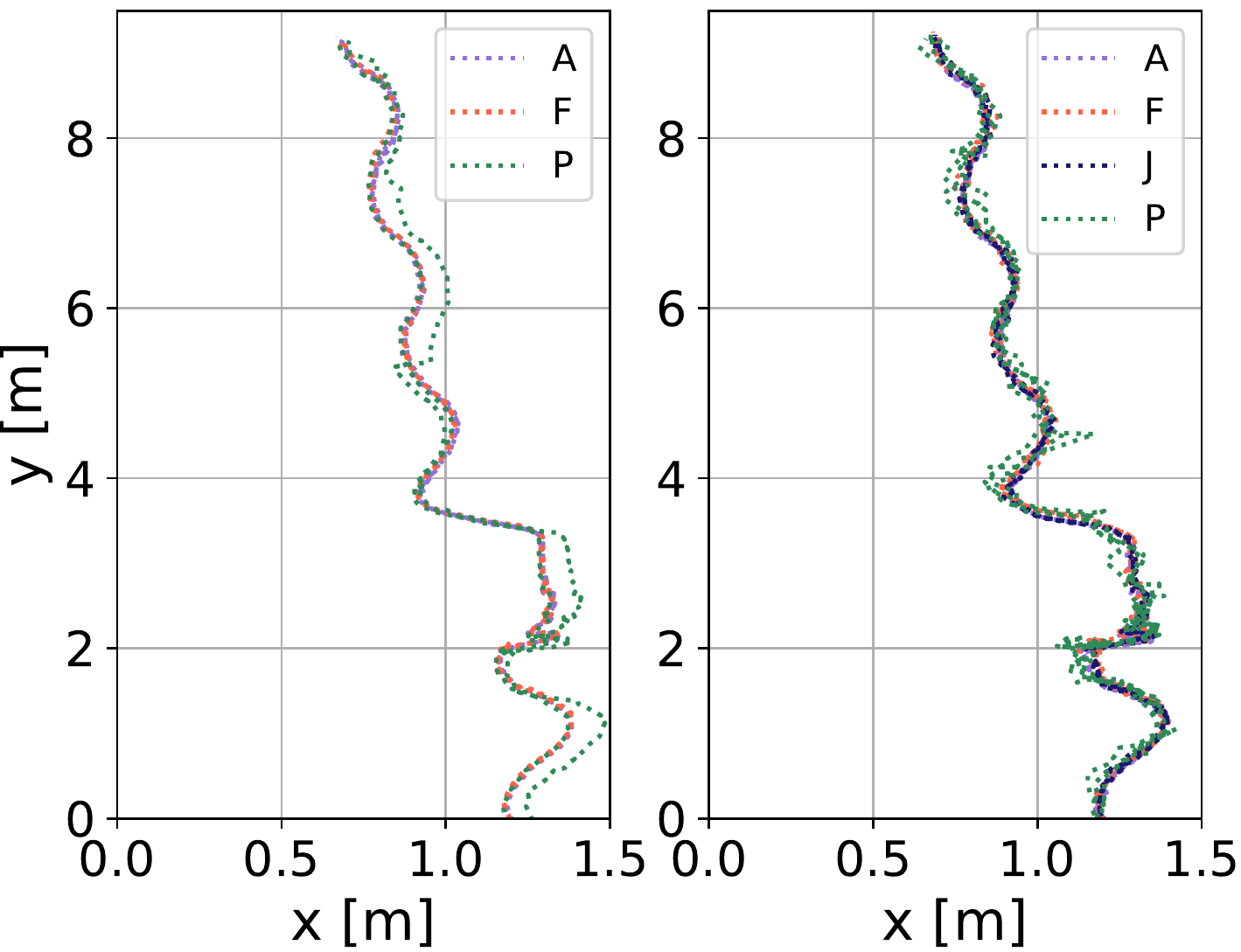}
        \caption{Left: 7 auto-tracked trajectories; Right: 12 manually tracked trajectories}
    \end{subfigure}
    \newline
    \begin{subfigure}{0.8\linewidth}
        \centering
        \includegraphics[width=\linewidth]{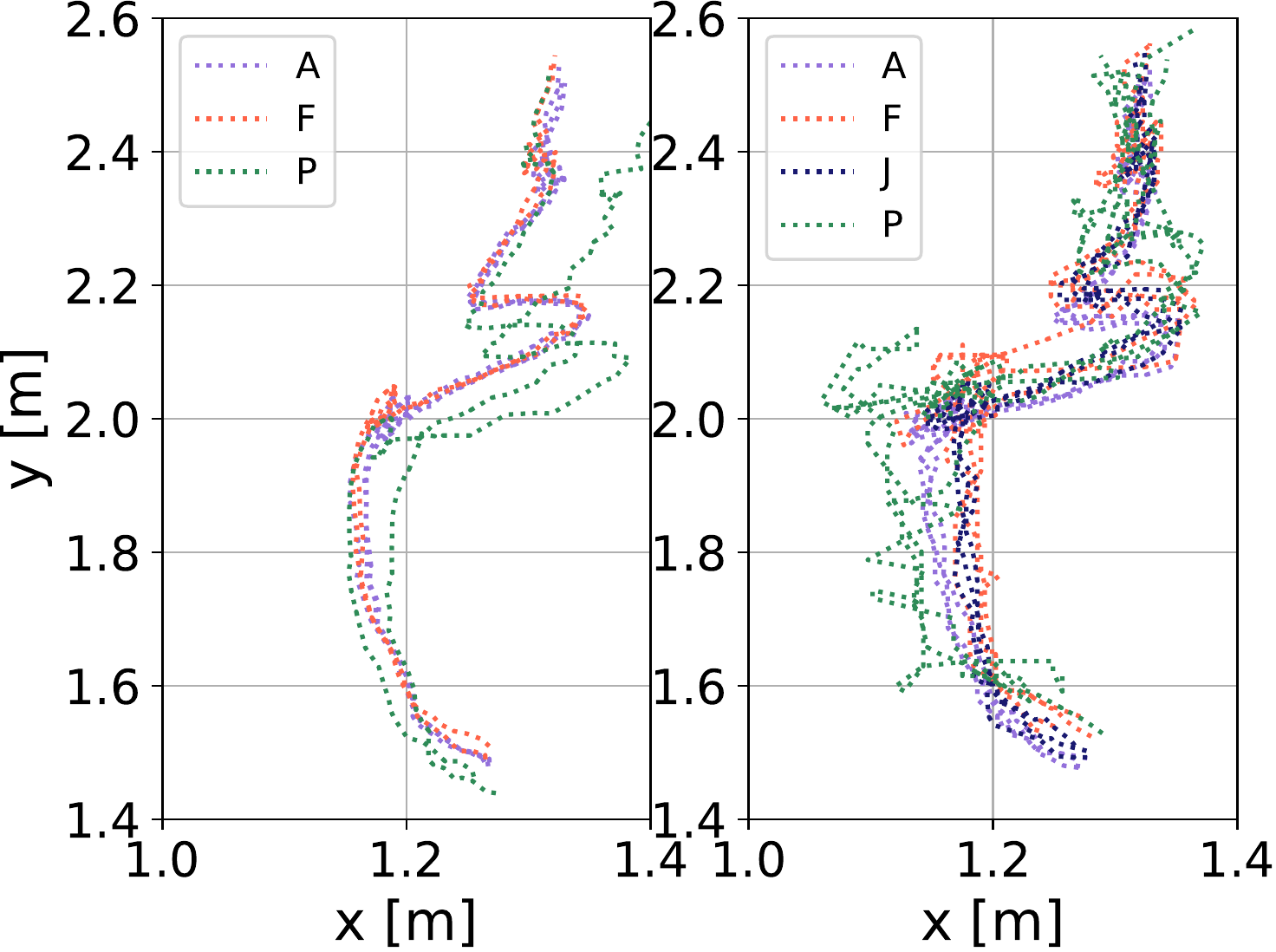}
        \caption{The same trajectories but zoomed in. The zoomed version shows a greater variance in the manually tracked trajectories than in the automatically tracked ones.}
    \end{subfigure}
    \caption{An excerpt of trajectories which were used to derive the measurement error from.}
    \label{fig:ExperimentDenseCrowdMeasurementErrorZoomed}
\end{figure}

The detailed graphical analysis in \fref{fig:ExperimentDenseCrowdMeasurementErrorZoomed} and statistical analysis in \tref{tbl:ExperimentDenseCrowdMeasurementError} revealed that \program{Tracker}'s auto-tracking feature results in a smaller measurement error than manually tracking.

\begin{table}[!h]
    \begin{tabular}{lrr}
        \toprule
        {} &         \textbf{Auto} &  \textbf{Manual} \\
        {} &         \textbf{Tracking} &  \textbf{Tracking} \\
        {} &         \textbf{Error} &  \textbf{Error} \\
        {} &         \textbf{[\si{cm}]} &  \textbf{[\si{cm}]} \\
        \midrule
        Mean   &   3.88 &   4.83 \\
        Std    &   3.32 &   3.43 \\
        \midrule
        Sample size &  5628 &  17688 \\
        \bottomrule
    \end{tabular}
    \caption{The measurement error for the trajectory extraction is smaller when using \code{Tracker}'s auto-tracking feature instead of manually tracking a participant's trajectory.}
    \label{tbl:ExperimentDenseCrowdMeasurementError}
\end{table}

\section*{Acknowledgement}

We thank all participants for taking part in the experiment. Additionally, we thank the student team consisting of Karim Belkhiria, Fabian Flach, Jonas Goltz, André Heinrich, Alexandra Mayer, Maximilian Niedermaier, Amir Schnell, Philipp Schuegraf for assisting with the experiment and pre-analysing the video material. We thank Dr. Mohcine Chraibi (Forschungszentrum Jülich) for simulating the experiment setup with \program{JuPedSim}. Last but not least, we thank the research office (FORWIN) of the Munich University of Applied Sciences and the Faculty Graduate Center CeDoSIA of TUM Graduate School at Technical University of Munich for their support. The experiment was approved by the ethical review committee of Technical University Munich (\url{https://ek-med-muenchen.de/}) and we carefully obtained an informed consent of all participants. The participants were not exposed to any danger throughout the experiment.

\paragraph{Authors' contributions} B.\,K. designed and conducted the experiment, analysed the experiment data and drafted the article. G.\,K. gave feedback, assisted with the experiment, critically revised the article, gave final approval for publication and took care of the funding acquisition. J.\,D. hosted B.\,K. at the School of Psychology (University of Sussex) for three months which allowed B.\,K. to get familiar with psychology of crowds and collective actions. J.\,D. also critically reviewed the article from a psychological perspective.

\paragraph{Funding} B.\,K. is supported by the German Federal Ministry of Education and Research through the project OPMoPS to study organised pedestrian movement in public spaces (grant no. 13N14562). This work is financially supported through the Open Access Publication fund of the Munich University of Applied Sciences.

\paragraph{Data accessibility} We analysed the video material and extracted trajectories of all participants. These trajectories and scripts for analysis are placed in the \enquote{Pedestrian Dynamics Data Archive} for public access: \url{http://ped.fz-juelich.de/da/doku.php?id=motion_through_crowd} (\url{https://doi.org/10.34735/ped.2019.1})

\clearpage


\end{document}